\documentclass[12pt]{article}
\usepackage{latexsym}
\usepackage{amsmath}

\renewcommand{\epsilon}{\varepsilon}

\renewcommand{\baselinestretch}{1.7}

\usepackage{amsmath,bm}
\usepackage{amssymb}
\usepackage{xcolor}
\usepackage{amsthm}
\usepackage[ansinew]{inputenc}
\usepackage{graphicx}
\usepackage{epsfig}
\usepackage{dsfont}
\usepackage{bbm}
\usepackage{subfig}
\usepackage{url}
\usepackage{placeins}

\usepackage[authoryear]{natbib}

\newtheorem{satz}{Theorem}[section]
\newtheorem{example}{Example}[section]

\newtheorem{rem}[satz]{Remark}
\newtheorem{assumption}[satz]{Assumption}
\newtheorem{algorithm}[satz]{Algorithm}

\def\3{\ss}
\def\er{\mathbb{R}}

\def\en{\mathbb{N}}

\newcommand{\bea}{\begin{eqnarray*}}
\newcommand{\eea}{\end{eqnarray*}}
\newcommand{\be}{\begin{eqnarray}}
\newcommand{\ee}{\end{eqnarray}}

\newcommand{\ba}{\begin{array}}
\newcommand{\ea}{\end{array}}

\def\3{\ss}
\def\er{\mathbb{R}}

\def\en{\mathbb{N}}

\newcommand{\Zc}{\mathcal{Z}}

\newcommand{\dn}{\stackrel{\cal D}{\longrightarrow}}

\newcommand{\Ec}{\mathcal{E}}
\newcommand{\Xc}{\mathcal{X}}

\usepackage[authoryear]{natbib}

\begin{document}
\def\spacingset#1{\renewcommand{\baselinestretch}%
{#1}\small\normalsize} \spacingset{1}

\title{{\bf Equivalence  of regression curves}}

  \author{
{\small  Holger Dette, Kathrin M\"ollenhoff,}\\
{\small Ruhr-Universit\"at Bochum }\\
{\small Fakult\"at f\"ur Mathematik} \\
{\small 44780 Bochum, Germany }\\
%{\small e-mail: holger.dette@rub.de}\\
\and
{\small Stanislav Volgushev }\\
{\small  Cornell University}\\
{\small   Department of Statistical Science} \\
{\small 301 Malott Hall  } \\
{\small Ithaca, NY 14853}\\
%{\small e-mail:  sv395@cornell.edu}\\
\and
{\small Frank Bretz}\\
{\small  Novartis Pharma AG}\\
{\small   4002 Basel, Switzerland} \\
{\small   and} \\
{\small Shanghai University of Finance and Economics}\\
{\small Shanghai 200433, China}\\
%{\small e-mail:   frank.bretz@novartis.com}\\
}

\pdfminorversion=4
 \maketitle

\begin{abstract}
This paper investigates the problem whether the difference between two parametric models $m_1, m_2$ describing the relation between a response variable and several covariates in two different groups is practically irrelevant, such that inference can be performed on the basis of the pooled sample. Statistical methodology is developed to test the hypotheses $H_0: d(m_1,m_2) \geq \varepsilon$ versus  $H_1: d(m_1,m_2) < \varepsilon$ to demonstrate equivalence between the two regression curves $m_1, m_2$ for a pre-specified threshold $\varepsilon$, where $d$ denotes a distance measuring the distance between $m_1$ and $m_2$. Our approach is based on the asymptotic properties of a suitable estimator $d(\hat m_1, \hat m_2)$ of this distance.
In order  to improve the approximation of the nominal level for small sample sizes a bootstrap test is developed, which addresses the specific form of the interval hypotheses.
In particular, data has to be generated under the null hypothesis, which implicitly defines a manifold for the parameter vector.
The results are illustrated by means of a simulation study and a data example. It is demonstrated that the new methods substantially improve currently available approaches with respect to power and approximation of the nominal level.
\end{abstract}
\vskip-.2cm
% \noindent AMS Subject Classification: 62P10, 62F03, 60G15 \\

\parindent 0cm

\spacingset{1.45}
\section{Introduction} \label{sec1}
\def\theequation{1.\arabic{equation}}
\setcounter{equation}{0}

Testing statistical hypotheses of equivalence has grown significantly in importance over the last decades, with applications covering such different areas as comparative bioequivalence trials, evaluating negligible trend in animal population growth, and model validation; see, for example, \cite{cade2011} and the references therein. Equivalence tests are based on a null hypothesis that a parameter of interest, such as the effect difference of two treatments, is outside an equivalence region defined through an appropriate choice of an equivalence threshold, denoted as $\varepsilon$ in this paper. If the null hypothesis is rejected one can then claim at a pre-specified significance level $\alpha$ that, in the previous example, the two treatments have an equivalent effect [see \cite{wellek2010testing}]. Equivalence testing is often used in regulatory settings because it reverses the burden of proof compared to a standard test of significance.

In this paper, we consider the problem of establishing equivalence of two
regression models which are used for the description of the relation between
a response variable and several covariates for two different groups, respectively.
That is, the objective is to
investigate whether the difference between these two models for the two groups is practically irrelevant,
so that only one  model can be used for both groups based on the pooled sample.
Such problems appear for example in population pharmacokinetics (PK) where the goal is to establish bioequivalence
of the concentration profiles over time, say $m_1$,  $m_2$, of two compounds. Traditionally,
bioequivalence is established by demonstrating equivalence between real valued quantities such as the
area under the curve (AUC) or the maximum concentrations ($C_{\max}$) [see \cite{chowliu1992,haustepig2007}].
However, such an approach may be misleading because the two profiles could be very different
although they may have similar AUC or $C_{\max}$ values. Hence it  might be more reasonable
to work directly with the underlying PK profiles instead of the derived summary statistics.

Another application of comparing two dose response curves occurs when assessing the
results from one patient population relative to another. For example, the international 
regulatory guidance document \cite{ich1997}
describes the concept of a bridging study based on, for example, the request of a new geographic region to determine 
whether data from another region are applicable to its population. If the bridging study shows that
dose response, safety and efficacy in the new region are similar to another region, then the study is
readily interpreted as capable of bridging the foreign data.
As a result, the ability of extrapolating foreign data to a new region
depends upon the similarity between the two regions.
The ICH E5 guidance does not provide a precise definition of similarity
and various concepts have been used in the literature.
For example, \cite{tsouetal2011} proposed a consistency approach for the  assessment of similarity
between a bridging study conducted in a new region and studies
conducted in the original region.
On the other hand, the ICH E5 guidance does require that
the safety and efficacy profile in the new region is not substantially different from
that in the original region, and
similarity can therefore be  interpreted as demonstrating ``no substantial difference'', 
which results in an equivalence testing problem [see \cite{liuhsuchen2002}].

The problem of establishing  equivalence of two regression models while controlling the Type I error rate has found considerable attention in the recent literature. For example,  \cite{liubrehaywynn2009}
proposed tests for the hypothesis of equivalence of two regression functions, which are applicable in linear models. \cite{gsteiger2011}
considered non-linear models and suggested a bootstrap method which is based on a  confidence band for the difference of the two regression models
[see also \cite{liuhaywyn2007}].
Both references use the intersection-union principle  [see for example \cite{berger1982}] to construct an overall  test for equivalence. We demonstrate in this paper that this approach leads to rather conservative test procedures with low power. Instead, we propose to directly estimate the distance, say $d(m_1,m_2)$, between the regression curves $m_1$ and $m_2$ and to decide for the
equivalence of the two curves if the estimator is smaller than a given threshold. The critical values of this test can be obtained by asymptotic theory, which describes the limit
distribution of an appropriately standardized estimated distance. In order to improve the approximation of the nominal level for small samples sizes a   non-standard bootstrap approach is proposed.

In
Section \ref{sec2} we introduce the general problem of demonstrating the equivalence between two regression curves.
While the concept of similarity of the two profiles is formulated  for a general distance $d$, we concentrate in the subsequent
discussion on two specific cases.  Section \ref{sec3} is devoted to
the comparison of curves with respect to $L^2$-distances.  We prove  asymptotic normality of the corresponding test statistic and construct an asymptotic level-$\alpha$ test. Moreover,  a non-standard bootstrap procedure is introduced, which addresses the particular difficulties arising in the problem of testing (parametric) interval hypotheses. In particular, resampling has to be performed under the null hypothesis $H_0: d(m_1,m_2)\geq \epsilon$, which defines (implicitly) a manifold in the parameter space. We prove consistency of the bootstrap test and demonstrate by means of a simulation study that it yields  an improvement of the approximation of the nominal level for small sample sizes. In Section \ref{sec4} the maximal deviation between the two curves is considered as a measure of similarity, for which corresponding results are substantially harder to derive. For example, we prove weak convergence of a corresponding test statistic, but the limit distribution depends in a complicated way on the extremal points of the  difference between the ``true'' curves. This problem is again solved by developing a bootstrap test.
The finite sample properties of the new methodology are illustrated in Section \ref{sec5}, where we also provide a comparison with the method of \cite{gsteiger2011}. In particular, it is demonstrated that the methodology proposed in this paper is more powerful than the test proposed by these authors.
The methods are illustrated with an example in Section \ref{sec6}. Technical details and proofs are deferred to an appendix in Section \ref{sec7}.

\section{Equivalence of  regression curves} \label{sec2}
\def\theequation{2.\arabic{equation}}
\setcounter{equation}{0}

We consider two, possibly different, regression models $m_1$,  $m_2$ to  describe the relationship between
a response variable $Y$ and several covariates for two different groups $\ell =1,2$:
 \be \label{mod1a}
 Y_{\ell ,i,j} &=& m_\ell (x_{\ell ,i},\beta_\ell )+\eta_{\ell ,i,j}~,~j=1,\ldots , n_{\ell ,i},~i=1, \ldots, k_\ell .
%   Y_{2,i,j} &=& m_2(x_{2,i},\beta_2)+\eta_{2,i,j}  ~,~j=1,\ldots , n_{2,i},~i=1, \ldots, k_2.  \label{mod1b}
  \ee
Here, the  covariate region is denoted by $\mathcal{X}\subset \mathbb{R}^d$,
$x_{\ell, i} $ denotes  the \emph{i}th dose level  (in group $\ell$), $n_{\ell,i}$  the number of patients treated  at dose level $x_{\ell, i} $ and $k_\ell$ the  number of different dose levels in group $\ell$.  Further, $n_\ell =\sum_{i=1}^{k_\ell}n_{\ell ,i}$ denotes
the sample size in group $\ell$, $n=n_1+n_2$ the total sample size,
and  the functions  $m_1$ and $m_2$ in \eqref{mod1a}
%and  \eqref{mod1b}
define the (non-linear)  regression models with $p_1$-  and $p_2$-dimensional
parameters  $\beta_1$ and $\beta_2$, respectively. The error terms are
assumed to be independent and identically  distributed with mean $0$ and variance $\sigma_\ell^2$ for group $\ell =1,2$.
Let  $({\cal M},d)$ denote a metric space of real valued functions of the form  $g: \mathcal{X} \to \er$  with  distance $d$. We
assume for all   $\beta_1$, $\beta_2$ that the regression functions satisfy $m_1(\cdot,\beta_1), m_2(\cdot,\beta_2) \in {\cal M} $, identify the models $m_\ell$ by their parameters $\beta_\ell$ and denote the distance between the two models by $d(\beta_1,\beta_2)(=d(m_1,m_2))$.\\
We consider the curves $m_1$ and $m_2$ as equivalent
if the distance between the two curves is small, that is $d(\beta_1,\beta_2)  < \epsilon $, where $\varepsilon$
is a pre-specified positive constant. In clinical practice, $\varepsilon$ is often denoted as relevance threshold in the sense that if
$d(\beta_1,\beta_2)  < \epsilon $ the difference between the two curved is believed not to be clinically relevant.
   In order to establish equivalence of the two dose response curves, we formulate the hypotheses
 \be H_0:  d(\beta_1,\beta_2)  \ge  \epsilon  \quad \mbox{  versus } \quad H_1:  d(\beta_1,\beta_2)  < \epsilon,
\label{hypgen}
\ee
which in the literature are called precise hypotheses, following \cite{berger1987}. 
%The choice of the distance $d$ depends on the specific problem under consideration.
The choice of $\varepsilon$ depends on the particular problem under consideration. 
For example, when testing for bioequivalence we can conclude that two treatments are not different from one another if the 90\% 
confidence interval of the ratio of a log-transformed exposure measure (AUC and/or $C_{\max}$; see Section~\ref{sec1}) 
falls completely within the range 80-125\%, indicating that differences in systemic drug exposure within these limits 
are not clinically significant [see \cite{food2003guidance}].
For the comparison of  dissolution profiles, which is a special case of the problem considered
in this paper, we refer to Appendix I of \cite{ema} with some recommendations
for the choice of the equivalence threshold on the basis of
univariate measures [see for example \cite{yukkanbay2000}].

In the following we are particularly interested in the metric space of all continuous functions with  distances
 \be
  \label{sup}
 d_\infty (\beta_1,\beta_2)  &=&  \max_{x\in \mathcal{X}} |m_1(x,\beta_1)-m_2(x,\beta_2)| \\
% \ee
% and of all square integrable functions with  distance
% \be
  \label{ell2}
  d_2 (\beta_1,\beta_2)  &=&   \int_\mathcal{X}  | m_1(x,\beta_1)-m_2(x,\beta_2) | ^2 dx.
 \ee
The maximal deviation distance $d_\infty$ is of interest, for example, in drug stability studies, where one investigates whether the maximum difference in mean drug content between two batches is no  larger than a pre-specified threshold; see, for example, \cite{rubhsu1992} and \cite{liuetal2007}. The $L^2$-distance $d_2$ might be attractive for demonstrating similarity of, for example, two PK models because  it  measures the squared integral of the difference between the two curves and is therefore related to the areas under the curves, which in turn is often of interest in bioequivalence studies, as mentioned above.
\\
The maximal deviation distance \eqref{sup} has also been considered in \cite{liubrehaywynn2009} and \cite{gsteiger2011}, who constructed confidence bands for the difference of two regression curves and used the intersection-union principle to derive an overall test for the hypothesis
that the two curves are equivalent.  In  linear models with normally distributed errors this test keeps the significance level not only asymptotically, but exactly at level $\alpha$ for any fixed sample size [see also,  \cite{bharspur2004}  or
 \cite{liulinpie2008} for some exact confidence bounds when comparing two linear regression models].
However, the resulting test turns out to be conservative and has low power, as demonstrated in Section \ref{sec5}.
This observation can be explained by the fact that the ``classical''
inversion of a confidence interval for a parameter, say $\mu$,  provides a level $\alpha$-test for the hypothesis $H_0: \mu = 0$,
but  it  yields usually  a  conservative test for the hypothesis
 $H_0: | \mu | \geq \varepsilon$ [see \cite{wellek2010testing}]. The same phenomenon also appears in the present context of comparing curves.
These properties  may limit the  use of the procedures proposed by  \cite{liubrehaywynn2009} and \cite{gsteiger2011}
in practice, as we would
like to maximize the probability of rejecting the null hypothesis if the two regression curves are in fact equivalent as measured by the relevance threshold $\varepsilon$.
\\
In the following, we develop alternatives approaches that are  more powerful. Roughly speaking,
we consider for $\ell = 1,2$ the estimator $  m_\ell (\cdot, \hat \beta_\ell)$ of the regression curve $m_\ell$ and reject the null hypothesis \eqref{hypgen}
for small values of the statistic $\hat d=d(\hat \beta_1, \hat \beta_2)$. The critical values can be obtained
by asymptotic theory deriving the limit distribution of $\sqrt{n} \ (\hat d-d)$ if  $n_1,n_2 \to \infty$, as developed in the following sections. This approach leads to a satisfactory solution  for the $L^2$-distance \eqref{ell2} based on the quantiles of the normal distribution (see Section \ref{sec3}). However, for the maximal deviation distance \eqref{sup}, the limit distribution depends in a complicated way on the extremal points
$$\mathcal{E} = \{ x \in \mathcal{X} \mid |\Delta (x,\beta_1,\beta_2)| = d_\infty(\beta_1,\beta_2)\}$$
of the true difference
 \begin{align} \label{truediff}
 \Delta(x,\beta_1,\beta_2) =m_1(x,\beta_1)-m_2(x,\beta_2).
 \end{align}
Moreover, in small sample  trials  the approximation of the nominal level of a given test based on asymptotic theory may not be valid. In order to obtain a more accurate approximation of the nominal level, we propose a non-standard bootstrap procedure and prove its consistency.
This procedure has to be constructed in a way such that it addresses the particular features of the equivalence hypotheses \eqref{hypgen}.
In particular, data have to be generated under the null hypothesis $d(\beta_1, \beta_2) \geq \varepsilon$, which implicitly defines a manifold for the vector of parameters $(\beta^T_1, \beta^T_2)^T\in \er^{p_1+p_2} $ of both models.
The non-differentiability of the maximal deviation distance $d_\infty$ exhibits some technical difficulties of such an approach, and for this reason we
begin the discussion with the $L^2$-distance $d_2$.

\section{Comparing curves by $L^2$-distances} \label{sec3}
\def\theequation{3.\arabic{equation}}
\setcounter{equation}{0}

In this section we construct a test for the equivalence of the two regression curves with respect to
the squared  $L^2-$distance, i.e. we consider hypotheses of the form
\begin{align}  \label{hypothesesL2}
H_0:  \int_\mathcal{X}(m_1(x,\beta_1)-m_2(x,\beta_2))^2 dx\geq \epsilon_2\mbox{~~versus~~}
H_1: \int_\mathcal{X}(m_1(x,\beta_1)-m_2(x,\beta_2))^2 dx< \epsilon_2.
\end{align}
Note  that under certain regularity assumptions (see the Appendix for details) the ordinary least squares (OLS)
estimators, say $\hat \beta_1$ and  $\hat \beta_2$, of the parameters ${\beta_1}$ and  ${\beta_2}$
can usually be linearized in the form
  \begin{equation} \label{exp}
  \sqrt{n_\ell} \ (\hat \beta_\ell - \beta_\ell) = \frac {1}{\sqrt{n_\ell}} \sum^{k_\ell}_{i=1} \sum^{n_{\ell,i}}_{j=1} \phi_{\ell,i,j}+ o_{\mathbb{P}}(1), \qquad \ell=1,2,
  \end{equation}
  where the functions $\phi_{\ell,i,j}$ are given by
  \begin{equation} \label{phij}
  \phi_{\ell,i,j} = \tfrac {\eta_{\ell,i,j}}{\sigma^2_\ell}\Sigma^{-1}_\ell  \tfrac {\partial}{\partial b_\ell}  m_\ell (x_{\ell,i,}, b_\ell )\big|_{b_\ell=\beta_\ell}, \qquad \ell=1,2,
  \end{equation}
    and  the $(p_\ell \times p_\ell)-$dimensional  matrices $\Sigma_\ell$ are defined by
\begin{equation}\label{sigmal}
\Sigma_\ell ={1 \over \sigma_\ell ^2} \sum_{i=1}^{k_\ell}\zeta_{\ell,i}\tfrac {\partial}{\partial b_\ell}  m_\ell (x_{\ell,i,}, b_\ell )\big|_{b_\ell=\beta_\ell}
 \big(\tfrac {\partial}{\partial b_\ell}  m_\ell (x_{\ell,i,}, b_\ell )\big|_{b_\ell=\beta_\ell} \big)^T~,~~\ell=1,2.
\end{equation}
For these arguments  we assume that
the matrices $\Sigma_\ell$ are non-singular and that
the sample sizes $n_\ell$ converge to infinity such that
\be
\label{des}
\lim_{n_\ell \to \infty }  {n_{\ell,i} \over n_\ell} = \zeta_{\ell,i} > 0 ~,~~i=1 , \ldots , k_{\ell}~,~\ell=1,2,
\ee
and
\begin{equation}\label{lambda}
\lim_{n_1,n_2\rightarrow \infty}\frac{n}{n_1} =  \lambda\in (1,\infty).
\end{equation}
It then follows by  straightforward calculation that the OLS estimators
are asymptotically normal distributed, i.e.
\be \label{MLasy}
&& \sqrt{n_\ell }(\hat \beta_\ell -\beta_\ell )\stackrel{\mathcal{D}}{\rightarrow}\mathcal{N}(0,\Sigma_\ell ^{-1})~,~\ell=1,2,
\ee
where the symbol $\dn$ means  weak convergence (convergence in distribution for real valued random variables).
The asymptotic variance in \eqref{MLasy} can easily be estimated   by replacing the parameters $\beta_\ell$, $\sigma_\ell$ and $\zeta_{\ell,i}$
in \eqref{sigmal}   by their estimators $\hat \beta_\ell$, $\hat \sigma_\ell$ and $n_{\ell,i}/n_\ell \ (\ell=1,2)$. The resulting estimator will be denoted by
$\hat \Sigma_\ell $ throughout this paper.
The null hypothesis in \eqref{hypothesesL2} is then rejected whenever
\begin{equation}
\hat d_2 :=
d_2 ( \hat \beta_1,\hat \beta_2)= \int_\mathcal{X}(m_1(x,\hat{\beta_1})-m_2(x,\hat{\beta_2}))^2 dx<c\label{testL2},
\end{equation}
where $c$ denotes a pre-specified constant defined through the level of the test. In order to determine this constant
  we will   derive the asymptotic distribution
of the statistic $\hat d_2$. The following result  is proved in the Appendix.  \smallskip

\begin{satz} \label{thm1}
If Assumptions \ref{2.0} - \ref{2.5} from the Appendix, \eqref{des} and \eqref{lambda} are satisfied,
 we have
\begin{equation}\sqrt{n} {(\hat{d}_2- d_2)}\stackrel{\mathcal{D}}{\longrightarrow}\mathcal{N}(0,{\sigma_{d_2}^2}),
\label{norm1}\end{equation}
where the asymptotic variance is given by
\begin{eqnarray}
\sigma_{d_2}^2=\sigma^2_{d_2}(\beta_1,\beta_2)=4 {\int_{\mathcal{X}\times\mathcal{X}}\Delta(x,\beta_1,\beta_2) \Delta(y,\beta_1,\beta_2)  k(x,y)dx dy},
\label{sigmat}
\end{eqnarray}
$\Delta(x,\beta_1,\beta_2)$ is defined in  \eqref{truediff}
and the kernel $k(x,y)$ is given by
\begin{eqnarray}
k(x,y)&:=& \lambda
\big(\tfrac {\partial}{\partial b_1}  m_1 (x, b_1 )\big|_{b_1=\beta_1}\big)^T \Sigma_1^{-1}\tfrac {\partial}{\partial b_1}  m_1 (y, b_1 )\big|_{b_1=\beta_1}
\nonumber\\
 &+&\tfrac{\lambda}{\lambda-1} \big(\tfrac {\partial}{\partial b_2}  m_2 (x, b_2 )\big|_{b_2=\beta_2}\big)^T\Sigma_2^{-1}
\tfrac {\partial}{\partial b_2}  m_2 (y, b_2 )\big|_{b_2=\beta_2}.
\label{kernel}
\end{eqnarray}
\end{satz}

\smallskip

Theorem \ref{thm1} provides a simple asymptotic level-$\alpha$ test for the hypothesis \eqref{hypothesesL2} of equivalence of two regression curves. More precisely, if $\hat \sigma_{d_2}^2=\sigma^2_{d_2}(\hat \beta_1, \hat \beta_2)$ denotes the (canonical)
estimator of the asymptotic variance in \eqref{sigmat}, then the null hypothesis in \eqref{hypothesesL2} is rejected
if
\be \label{testl2asy}
\hat d_2 < \varepsilon_2 + \frac {\hat \sigma_{d_2}}{\sqrt{n}} u_\alpha,
\ee
where $u_\alpha$ denotes the $\alpha$-quantile of the standard normal distribution.  Note
that by the nature of the  problem the quantile of this  test  depends
on the threshold $\varepsilon_2$.
The finite sample properties of this test will be investigated in Section \ref{sec51}.

\bigskip

\begin{rem} \label{rem1}
{\rm
It follows from Theorem \ref{thm1} that the  test \eqref{testl2asy} has asymptotic level $\alpha$ and is consistent if $n_1,n_2 \to \infty$.
More precisely, if $\Phi$ denotes the cumulative distribution function of the standard normal distribution, we have
for the probability of rejecting the null hypothesis in \eqref{hypothesesL2}
\begin{eqnarray*}
 \mathbb{P} \Big( \hat d_2 < \varepsilon_2 + \frac {\hat \sigma_{d_2}}{\sqrt{n}} u_\alpha \Big) &=&
 \mathbb{P} \Big( \frac {\sqrt{n}}{\hat \sigma_{d_2}} (\hat d_2-d_2) < \frac {\sqrt{n}}{\hat \sigma_{d_2}} (\varepsilon_2 - d_2) + u_\alpha \Big) ~.
 %\\  & \leq & \mathbb{P} \Big( \frac {\sqrt{n_1+n_2}}{\hat \sigma_{d_2}} (\hat d_2-d_2) \leq u_\alpha\Big) \approx \Phi (u_\alpha)= \alpha,
\end{eqnarray*}
Under continuity assumptions it follows that  $\hat \sigma^2_{d_2} \stackrel{\mathbb{P}}{\longrightarrow} \sigma^2_{d_2}$ and   Theorem \ref{thm1} yields $\sqrt{n} \ (\hat d_2 - d_2)/ \hat \sigma_{d_2} \dn \mathcal{N}(0,1)$. This   gives
\begin{eqnarray*}
\mathbb{P} \Big( \hat d_2 \leq \varepsilon_2 + \frac {\hat \sigma_{d_2}}{\sqrt{n}} u_\alpha \Big)
\stackrel{\longrightarrow}{\scriptstyle {n_1,n_2 \to \infty }} \left \{
\begin{array}{ccc}
0 & \mbox{if} & d_2 > \varepsilon_2 \\
\alpha & \mbox{if}& d_2 =\varepsilon_2  \\
1 & \mbox{if} &d_2 < \varepsilon_2 \\
\end{array}
\right. \qquad .
\end{eqnarray*}
}
\end{rem}

The test \eqref{testl2asy} can be recommended if the sample sizes are reasonable large. However, we will demonstrate in Section \ref{sec5} that for very
small sample sizes, the critical values provided by this asymptotic theory may not provide an accurate approximation of the nominal level,
and for this reason we will also investigate a parametric bootstrap procedure to generate critical values for the statistic $\hat d_2$.

\medskip

\begin{algorithm} \label{alg1} {\rm (parametric bootstrap for testing precise hypotheses)
 \begin{itemize}
\item[(1)]
  Calculate the OLS-estimators $\hat{\beta_1}$ and $\hat{\beta_2}$, the corresponding variance estimators

  $$
  \hat \sigma^2_\ell = \frac {1}{n_\ell} \sum^{k_\ell}_{i=1} \sum^{n_{\ell,i}}_{j=1} (Y_{\ell,i,j}- m_\ell (x_{\ell,i}, \hat \beta_\ell ))^2; \qquad \ell=1,2,
  $$  and the test statistic
$\hat d_2=d_2 (\hat \beta_1, \hat \beta_2) $ defined by \eqref{testL2}.
\item[(2)] Define estimators of the parameters $\beta_1$ and $\beta_2$ by
\begin{equation} \label{MLcons}
{\hat{\hat{\beta}}_{\ell}}= \left\{
\begin{array} {ccc}
\hat \beta_\ell & \mbox{if} & \hat d_2 \geq \varepsilon_2 \\
\tilde \beta_\ell & \mbox{if} & \hat d_2 < \varepsilon_2
\end{array}  \right. \quad \ell=1,2,
\end{equation}
where $\tilde \beta_1, \tilde \beta_2$ denote the OLS-estimators of the parameters $\beta_1, \beta_2$
 under the constraint
	\begin{equation}\label{constr}
d_2(\beta_1,\beta_2)=\int_\mathcal{X}(m_1(x,\beta_1)-m_2(x,\beta_2))^2 dx = \epsilon_2.
\end{equation}
	Finally, define ${\hat{\hat{d}}_{2}} = d_2 ({\hat{\hat{\beta}}_{1}}, {\hat{\hat{\beta}}_{2}})$ and note that ${\hat{\hat{d}}_{2}}\geq \epsilon_2$.
\item[(3)] Bootstrap test
\begin{itemize}
\item[(i)]Generate bootstrap data under the null hypothesis, that is
	\begin{equation}\label{bootdata}
	Y_{\ell,i,j}^*=m_\ell(x_{\ell,i},\hat {\hat{\beta_\ell}})+\eta_{\ell,i,j}^* ~,i=1,\ldots, n_{\ell,i},~\ell=1,2 ,
	\end{equation}
	where the errors  $\eta^*_{\ell,i,j}$ are independent  normally distributed such that
	$\eta_{\ell,i,j}^* \sim \mathcal{N}(0, \hat  \sigma_\ell^2)$.
\item[(ii)] Calculate  the OLS estimators  $\hat \beta^*_1$  and $\hat \beta^*_2 $ and the test statistic
    $$\hat d^*_2=d_2(\hat \beta^*_1, \hat \beta^*_2)=\int_\mathcal{X}(m_1(x,\hat \beta^*_1)-m_2(x,\hat \beta_2^*))^2 dx$$
from the bootstrap data. Denote by $\hat{q}_{\alpha,2}$ the $\alpha-$quantile of the distribution of the statistic $\hat d^*_2$, which depends on the data $\left\{Y_{l,i,j}|l=1,2;\ j=1,...n_{l,i};\ i=1,...,k_l\right\}$ through the estimators $\hat{\hat {\beta}}_1$ and $\hat{\hat{\beta}}_2$.
\end{itemize}
The steps (i) and (ii) are repeated $B$ times to generate  replicates $\hat d^*_{2,1}, \dots, \hat d^*_{2,B}$ of $\hat d^*_2$.    If $\hat d^{*(1)}_2 \le \ldots  \le \hat d^{*(B)}_2$
denotes the corresponding order statistic, the  estimator of the  quantile of the distribution of $\hat d^*_2$
is defined by  $\hat{q}_{\alpha,2}^{(B)} :=  \hat d_2^{*(\lfloor B \alpha \rfloor )}$, and the null hypothesis is rejected  if
\begin{equation} \label{test}
\hat d_2 < \hat{q}_{\alpha,2}^{(B)}.
\end{equation}
Note that the bootstrap quantile $\hat{q}_{\alpha,2}^{(B)}$ depends on the threshold $\varepsilon_2$ which is used in
the hypothesis \eqref{hypothesesL2}, but we do not reflect this dependence in our notation.
\end{itemize}
}
\end{algorithm}

The following result shows that the bootstrap test \eqref{test} 
has asymptotic level $\alpha$ and is consistent if $n_1,n_2 \to \infty$. 
Its proof can be found in the Appendix.
%is a consistent asymptotic level-$\alpha$ test.

\begin{satz}   \label{thm2}
Assume that the conditions of Theorem \ref{thm1} are satisfied.
 \begin{itemize}
 \item[(1)] If the null hypothesis  in $\eqref{hypothesesL2}$  holds, then we have for any $\alpha \in (0, 0.5)$
\begin{equation}
\lim_{n_1,n_2 \rightarrow\infty}\mathbb{P} (\hat{d}_2<\hat{q}_{\alpha,2} )= \left\{
\begin{array} {ccc}
0 & \mbox{if} & d_2 > \varepsilon_2 \\
\alpha & \mbox{if} & d_2 = \varepsilon_2
\end{array}. \right.
\label{level}\end{equation}
 \item[(2)] If the alternative in $\eqref{hypothesesL2}$  holds, then   we have for any $\alpha \in (0, 0.5)$
\begin{equation}\lim_{n_1,n_2 \rightarrow\infty}\mathbb{P} (\hat{d}_2<\hat{q}_{\alpha,2} )=1.\label{consistence}\end{equation}
\end{itemize}
\end{satz}

\section{Comparing curves by their maximal deviation} \label{sec4}
\def\theequation{4.\arabic{equation}}
\setcounter{equation}{0}

In this section we construct a test for the equivalence of the two regression curves with respect to
the maximal absolute deviation  \eqref{sup}. The corresponding test statistic is given by the maximal deviation distance
\begin{equation}\label{dsup}
\hat d_\infty = d_\infty (\hat \beta_1, \hat \beta_2) = \max_{x \in \mathcal{X}} | m_1(x,\hat \beta_1) - m_2(x,\hat \beta_2) |
\end{equation}
between the two estimated regression functions,
where $\hat \beta_1, \hat \beta_2$ are the   OLS-estimators from the two samples.
In order to describe the asymptotic distribution of the statistic $\hat d_\infty$ we define the set of extremal points
\begin{equation} \label{eset}
\mathcal{E} = \big \{ x \in \mathcal{X}  \big | \ | m_1(x,\beta_1)-m_2(x,\beta_2)| = d_\infty \big \}
\end{equation}
and introduce the decomposition $\mathcal{E} = \mathcal{E}^+ \cup \mathcal{E}^-$,
where
\begin{equation} \label{esetpm}
\mathcal{E}^{\mp} = \big \{ x \in \mathcal{X} \big | \ m_1(x,\beta_1)-m_2(x,\beta_2)  = \mp \ d_\infty \big \}.
\end{equation}
The following result is proved in the Appendix.

\begin{satz}\label{thm3}
If $d_\infty >0$ and the assumptions of Theorem \ref{thm1} are satisfied, then
\begin{equation} \label{norm2}
\sqrt{n } \ (\hat d_\infty - d_\infty) \dn \mathcal{Z}:= \max \big \{  \max_{x \in \mathcal{E}^+} G(x),
\max_{x \in \mathcal{E}^-} (-G(x)) \big \},
\end{equation}
where $\{G(x)\}_{x \in \mathcal{X}}$ denotes  a Gaussian process defined by
\begin{equation}
G(x)=\big(\tfrac {\partial}{\partial b_1}  m_1 (x, b_1 )\big|_{b_1=\beta_1}\big)^T \sqrt{\lambda}\Sigma_1^{-1/2}
Z_1-\big(\tfrac {\partial}{\partial b_2}  m_2 (x, b_2 )\big|_{b_2=\beta_2}\big)^T\sqrt{\tfrac{\lambda}{\lambda-1}}
\Sigma_2^{-1/2} Z_2,
\label{G}
\end{equation}
and  $Z_1$ and $Z_2$ are  {independent} $p_1$- and $p_2$-dimensional  standard normal distributed random variables, respectively,
i.e. $Z_\ell \sim {\cal N} (0, I_{p_\ell})$, $\ell=1,2$.
\end{satz}

\smallskip

In principle, Theorem \ref{thm3} provides an asymptotic level $\alpha$-test for the hypotheses
\begin{equation} \label{h0inf}
H_0: d_\infty (\beta_1, \beta_2) \geq \varepsilon_\infty \quad \mbox{versus} \quad H_1: d_\infty (\beta_1, \beta_2) < \varepsilon_\infty
\end{equation}
by rejecting the null hypotheses whenever $\hat d_\infty < q_{\alpha,\infty}$, where $q_{\alpha,\infty}$ denotes the $\alpha$-quantile of the distribution of the random variable $\mathcal{Z}$ defined in \eqref{norm2}. However, this distribution has a very complicated structure. For example, if $ \mathcal{E}= \{x_0\}$ the distribution of $\mathcal{Z}$ is a centered normal distribution but with variance
\begin{eqnarray}\label{unique}
 \sigma_\infty^2&=&
  \lambda
\big(\tfrac {\partial}{\partial b_1}  m_1 (x_0, b_1 )\big|_{b_1=\beta_1}\big)^T \Sigma_1^{-1}\tfrac {\partial}{\partial b_1}  m_1 (x_0, b_1 )\big|_{b_1=\beta_1}\nonumber\\
&+& \tfrac{\lambda}{\lambda-1} \big(\tfrac {\partial}{\partial b_2}  m_2 (x_0, b_2 )\big|_{b_2=\beta_2}\big)^T\Sigma_2^{-1}
\tfrac {\partial}{\partial b_2}  m_2 (x_0, b_2 )\big|_{b_2=\beta_2}
\end{eqnarray}
 which depends on the location of the (unique) extremal point $x_0$. In general (more precisely in the case
 $\# \mathcal{E} >1$) the distribution of $\mathcal{Z}$ is the distribution of a maximum of dependent Gaussian random variables, where the variances and the dependence structure depend on the location of the extremal points of the function $\Delta(\cdot, \beta_1,\beta_2)$.
  Because the estimation of these points is very difficult, we propose a bootstrap approach to obtain suitable quantiles. The bootstrap test is defined in the same way as described in Algorithm \ref{alg1}, where the distance $d_2$ is replaced by the maximal deviation $d_\infty$. The corresponding quantile obtained in Step 3(ii) of Algorithm \ref{alg1} is now denoted by $\hat q_{\alpha,\infty}^{(B)}$, while the theoretical quantile of the bootstrap distribution is denoted by $\hat q_{\alpha,\infty}$. The following result is proved in the Appendix and shows that the test, which rejects the null hypothesis in \eqref{h0inf} whenever
\begin{equation} \label{testinf}
  \hat d_\infty < \hat q_{\alpha,\infty}^{(B)},
\end{equation}
has asymptotic level $\alpha$ and is consistent.
Interestingly the quality of the approximation of the nominal level of the test
depends on the cardinality of the set $\mathcal{E}$.

\begin{satz} \label{thm43}  Suppose that the assumptions of Theorem \ref{thm3}  hold.
\begin{itemize}
\item[(1)] If the null hypothesis in $\eqref{h0inf}$ is satisfied and the set $\mathcal{E}$ defined in \eqref{eset}  consists of one point, then we have for any $\alpha \in (0, 0.5)$
\begin{equation}\label{level2}
\lim_{n_1,n_2\rightarrow\infty}\mathbb{P}\big(\hat d_\infty<\hat q_{\alpha,\infty }\big) = \left\{
\begin{array} {ccc}
0 & \mbox{if} & d_\infty  > \varepsilon_\infty  \\
\alpha & \mbox{if} & d_\infty  = \varepsilon_\infty  .
\end{array} \right.
\end{equation}
\item[(2)] 
Let $F_\Zc$   denote the distribution function  of the random variable $\Zc$ defined in \eqref{norm2}  and   $q_{\Zc,\alpha}$ its $\alpha$-quantile. Assume that  $F_\Zc$ is continuous at $q_{\Zc,\alpha}$ and  $q_{\Zc,\alpha} < 0$. If the null hypothesis in $\eqref{h0inf}$ is satisfied 
we have 
\begin{equation}\label{level2.1}
\limsup_{n_1,n_2\rightarrow\infty}\mathbb{P}\big(\hat d_\infty<\hat q_{\alpha,\infty }\big)\leq\alpha.
\end{equation}
\item[(3)] If the alternative in $\eqref{h0inf}$  is satisfied, then we have for any $\alpha \in (0, 0.5)$
\begin{equation}
\lim_{n_1,n_2 \rightarrow\infty}\mathbb{P} \big(\hat d_\infty<\hat q_{\alpha,\infty }\big)=1.\label{consistence2}
\end{equation}
\end{itemize}
\end{satz}

\begin{rem} \label{rem41} ~~~ \\ {
\rm
(a)
The condition  in part (2) of Theorem \ref{thm43}  is a non-trivial assumption. By results in \cite{TS1976}, the distribution of $\Zc$ has at most one jump at the left boundary of its support and is continuous to the right of that. The condition on $F_\Zc$ to be continuous at $q_{\Zc,\alpha}$ is thus equivalent to requiring that the mass at the left endpoint of the support of $F_\Zc$ is smaller than $\alpha$. In some cases it is possible to show that $F_{\Zc}$ is continuous on $\mathbb{R}$, i.e. the mass at its left support point is zero. For example, this follows from Theorem 3 of \cite{CCK2015} provided that the condition
\begin{equation}\label{assumption_neu}
 \inf_{x\in\Ec} \Big( 
 \|\tfrac {\partial}{\partial b_1}  m_1 (x, b_1 )\big|_{b_1=\beta_1}\| + \|\tfrac {\partial}{\partial b_2}  m_2 (x, b_2 )\big|_{b_2=\beta_2}\| \Big)  > 0
\end{equation}
holds.  \\
The assumption \eqref{assumption_neu} is always fulfilled if one of the models  contains an additive (placebo) effect
because in this case  the first entry of the gradient $\tfrac {\partial}{\partial b_\ell}  m_\ell (x, b_\ell )\big|_{b_\ell=\beta_\ell}$ equals $1$.
Furthermore, if  the two models are of the form $m_\ell(x,\beta_\ell)=\beta_{\ell,1}\cdot m_\ell^0(x, \beta_\ell^0)$ for $l=1,2$
with 
$\beta_{\ell} = (\beta_{\ell,1},\beta_{\ell}^0)$
 (for example Michalis-Menten models), we have
$$\tfrac {\partial}{\partial b_\ell}  m_\ell (x, b_\ell )\big|_{b_\ell=\beta_\ell}=\big(m_\ell^0(x, \beta_\ell^0),\tfrac {\partial}{\partial b_\ell^0}  m_\ell^0 (x, b_\ell^0 )\big|_{b_\ell^0=\beta_\ell^0}\big),\ \ell=1,2.$$
Consequently, if \eqref{assumption_neu} was not fulfilled, there would exist $x_0\in\mathcal{E}$ such that $m_\ell^0 (x_0, \beta_\ell^0 )=0,\ \ell=1,2$,
 and as $x_0\in\mathcal{E}$ it holds $$d_\infty=\left|m_1(x_0,\beta_1)-m_2(x_0,\beta_2)\right|=\left|\beta_{1,1}\cdot m_1^0(x_0, \beta_1^0)-\beta_{2,1}\cdot m_2^0(x_0, \beta_2^0)\right|=0$$
This yields $m_1\equiv m_2$ and does not correspond to the null hypothesis.
\\
(c)
Note that the  asymptotic Type I error rate of the  bootstrap test is precisely $\alpha$  at the boundary of the hypothesis (i.e. $d_\infty  = \varepsilon_\infty$)
  if the cardinality of  $\mathcal{E}$ is one.  On the other hand, if  the set $\mathcal{E}$ contains more than one point, part (2) of Theorem \ref{thm43}
 indicates that the corresponding bootstrap  test is  usually  conservative, even at  the boundary of the hypothesis.  These results are confirmed
 by a simulation study in  Section \ref{sec52}.
 }
 \end{rem}

\section{Finite sample properties} \label{sec5}
\def\theequation{5.\arabic{equation}}
\setcounter{equation}{0}
\setcounter{table}{0}

In this section we investigate the finite sample properties of the asymptotic and bootstrap tests proposed
in Sections \ref{sec3} and \ref{sec4} in terms of power and size.
For the distance $d_\infty$ we also  provide a comparison with  the approach from \cite{gsteiger2011}.
Their method follows from \eqref{des} - \eqref{MLasy} and an
application of the Delta method [see for example \cite{vaart1998}]  so that the prediction
$m_1(x,\hat \beta_1)-m_2(x,\hat \beta_2)$ for the difference  of the two regression models at the point $x$ is approximately
normally distributed. That is,
 $$\frac{m_1(x,\hat \beta_1)-m_2(x,\hat \beta_2)-(m_1(x,\beta_1)-m_2(x,\beta_2))}{\hat \tau_{n_1,n_2}  (x,\hat \beta_1,\hat \beta_2)} \stackrel{\mathcal{D}}\longrightarrow \mathcal{N}(0,1),$$
 where
\be
\label{asyvar}
\hat \tau^2_{n_1,n_2}  (x, \hat \beta_1,\hat \beta_2) =
\sum_{\ell=1}^2 \tfrac{1}{n_\ell} \big( \tfrac{\partial}{\partial \beta_\ell}  m_\ell (x,\ \beta_\ell)\big|_{ \beta_\ell =\hat \beta_\ell} \big) ^T\hat \Sigma_\ell^{-1}
\tfrac{\partial}{\partial \beta_\ell} m_\ell (x, \beta_\ell)\big|_{ \beta_\ell =\hat \beta_\ell}
\ee
and $\hat \Sigma_\ell$ denotes the estimator of the variance in \eqref{sigmal}, which is obtained by replacing the parameters $\sigma_\ell$, $\beta_\ell$ and $\zeta_{\ell,i}$  by their estimators $\hat \sigma_\ell$, $\hat \beta_\ell$, and $n_{\ell,i}/n_\ell \ (\ell=1,2)$.  \cite{gsteiger2011} proposed a test based on the pointwise confidence bands derived by \cite{liuhaywyn2007}, that is
 \[m_1(x,\hat \beta_1)-m_2(x,\hat \beta_2)\pm\ z_{1-\alpha}   \hat  \tau_{n_1,n_2}  (x, \hat \beta_1,\hat \beta_2),  \]
where $z_{1-\alpha}$ denotes the $(1-\alpha)$-quantile of the standard normal distribution.
%where $z_{1-\alpha}$ is   the  $(1-\alpha)$-quantile of the distribution of the random variable
%$$
%D:=\max_{x\in \mathcal{X}}\frac{ |\frac{1}{\sqrt{n_1}}\tfrac{\partial}{\partial \beta_1} \big( m_1(x,\beta_1)\big|_{ \beta_1=\hat \beta_1} \big)^T \hat \Sigma_1^{-1/2}
%Z_1-\frac{1}{\sqrt{n_2}} \big(\tfrac{\partial}{\partial \beta_2} m_2(x,\beta_2)\big|_{ \beta_2 =\hat \beta_2} \big) ^T \hat \Sigma_2^{-1/2}  Z_2|}{\tau_{n_1,n_2}  (x, \hat \beta_1,\hat \beta_2) }
 %$$
 %and $Z_1$ and $Z_2$ are independent $p_1$- and $p_2$-dimensional standard normal distributed random variables, respectively. They proposed to determine this quantile by simulation, but they
 %did not prove that this parametric bootstrap method is in fact a valid procedure. On the other hand,  they demonstrated by means of a simulation study that the confidence bands obtained by this method  have   rather accurate
 %coverage probabilities.
A test for the hypotheses \eqref{h0inf}
is finally obtained  by rejecting the null hypothesis and conclude for equivalence, if the maximum (minimum) of the upper (lower) confidence band is smaller (larger)
than $ \epsilon_\infty$ ($- \epsilon_\infty$).  A particular advantage of this test is that it directly refers to the distance \eqref{sup}, which has a nice interpretation
in many applications. Moreover, in linear models (with normally distributed errors) it is an exact level-$\alpha$ test.
However, the resulting test is conservative and has low power compared to the methods proposed
in this paper as shown in Section \ref{sec52}.
%Note that it would also be possible to conduct the test by using the by \cite{gsteiger2011} proposed simultaneous confidence bands, which can be obtained by replacing the quantile of the normal distribution by a quantile of a parametric bootstrap approach. However, this test would be even less powerful.

All results in this and the following section are based on $1000$ simulation runs and the quantiles of the bootstrap tests have been obtained by $B=300$ bootstrap replications.  In all examples  the dose range is given by the interval ${\cal X}  =[0,4]$  and an equal number of patients is allocated
at the five dose levels  $x_{\ell,1}=0,x_{\ell,2}=1$, $ x_{\ell,3}=2$, $x_{\ell,4}=3$, $x_{\ell,5}=4$
in both groups (that is $k_1=k_2=5$).

\subsection{Tests based on the distance $d_2$} \label{sec51}

For the sake of brevity we restrict ourselves to a  comparison of two shifted EMAX-models

\be
\label{ex2}
m_1(x,\beta_1)= \beta_{11} + \frac{\beta_{12} x}{\beta_{13}+x}  \qquad \mbox{and} \qquad  m_2(x,\beta_2)= \beta_{21}+\frac{\beta_{22}x}{\beta_{23}+x},
\ee
where $\beta_{1} =  ( \beta_{11}, \beta_{12}, \beta_{13})  =(\delta, 5,1)$ and $\beta_{2} =  ( \beta_{21}, \beta_{22}, \beta_{23} ) =(0,5,1)$. 
In Tables \ref{tab5} and \ref{tab6} we  display the simulated Type I error rates of the bootstrap test
\eqref{test} and the asymptotic test \eqref{testl2asy} for $\epsilon_2 = 1$ in \eqref{hypothesesL2} and
various configurations of $\sigma^2_1$, $\sigma^2_2$, $n_1$, $n_2$ and $\delta$. In the interior of the null hypothesis (i.e. $d_2>\epsilon_2$)
the Type I error rates of the tests \eqref{testl2asy} and \eqref{test}  are   smaller than the nominal level  as predicted by
Remark \ref{rem1}.
     For both tests we   observe a rather  precise approximation of the nominal level (even for small sample sizes)
at the boundary of the  null hypothesis   (i.e. $d_2=1$). In some cases
the approximation of the nominal level by the bootstrap test  \eqref{test} is slightly more accurate and for this reason we recommend to use the bootstrap
test  \eqref{test} to establish equivalence of two regression models with respect to the $L^2$-distance. \\

%\FloatBarrier

 \begin{table}[!h]
  {\tiny
\centering
\begin{tabular}{||c|c|c|ccc||ccc||}
\hline
\multicolumn{1}{||c|}{} & \multicolumn{1}{c|}{} & \multicolumn{1}{|c|}{} & \multicolumn{3}{|c||}{$\alpha=0.05$} & \multicolumn{3}{|c||}{$\alpha=0.1$} \\\hline
\multicolumn{1}{||c|}{} & \multicolumn{1}{c|}{} & \multicolumn{1}{|c|}{} &
\multicolumn{3}{|c||}{$(\sigma_1^2,\sigma_2^2)$} & \multicolumn{3}{|c||}{$(\sigma_1^2,\sigma_2^2)$} \\\hline
$(n_1,n_2)$ & $\delta$ & $d_2$ & $(0.25,0.25)$ & $(0.5,0.5)$ & $(0.25,0.5)$ &
$(0.25,0.25)$ & $(0.5,0.5)$ & $(0.25,0.5)$
  \\ \hline
$(10,10)$ & 1 & 4       & 0.000 & 0.000 & 0.000 & 0.000 & 0.000 &0.000 \\
$(10,10)$ & 0.75 & 2.25 & 0.004 & 0.002 & 0.001 &  0.000 & 0.002 &0.000  \\
$(10,10)$ & 0.5 & 1     & 0.051 & 0.064 & 0.052 &  0.101 & 0.120 &0.118 \\
\hline
$ (10,20)$ & 1  & 4     & 0.000 & 0.000 & 0.000 &  0.000 & 0.000 &0.000 \\
$ (10,20)$ & 0.75 & 2.25& 0.000 & 0.000 & 0.000 &  0.000 & 0.000 &0.000  \\
$ (10,20)$ & 0.5 & 1    & 0.055 & 0.060 & 0.051 &   0.104 & 0.111 &0.101 \\
\hline
$ (20,20)$ & 1  & 4     & 0.000 & 0.000 & 0.000 & 0.000 & 0.000 &0.000 \\
$ (20,20)$ & 0.75 & 2.25& 0.001 & 0.002 & 0.000  &  0.004 & 0.005 &0.001 \\
$ (20,20)$ & 0.5 & 1    & 0.057 & 0.058 & 0.050   &  0.125 & 0.107 &0.097 \\
\hline
$(50,50)$ & 1 & 4       & 0.000 & 0.000 & 0.000  &  0.000 & 0.000 &0.000\\
$(50,50)$ & 0.75 & 2.25 & 0.001 & 0.000 & 0.000  & 0.002 & 0.000 &0.000 \\
$(50,50)$ & 0.5 & 1     & 0.057 & 0.048 & 0.054  &  0.097 & 0.114 &0.093 \\
\hline
\hline
\end{tabular}
\caption{\label{tab5} \it Simulated Type I error rates of the bootstrap test \eqref{test} for the equivalence of two shifted EMAX models defined in \eqref{ex2}. The threshold in \eqref{hypothesesL2} is chosen as $\epsilon_2=1$.}
}
\end{table}

\begin{table}[!h]
  {\tiny
\centering
\begin{tabular}{||c|c|c|ccc||ccc||}
\hline
\multicolumn{1}{||c|}{} & \multicolumn{1}{c|}{} & \multicolumn{1}{|c|}{} & \multicolumn{3}{|c||}{$\alpha=0.05$} & \multicolumn{3}{|c||}{$\alpha=0.1$} \\\hline
\multicolumn{1}{||c|}{} & \multicolumn{1}{c|}{} & \multicolumn{1}{|c|}{} &
\multicolumn{3}{|c||}{$(\sigma_1^2,\sigma_2^2)$} & \multicolumn{3}{|c||}{$(\sigma_1^2,\sigma_2^2)$} \\\hline
$(n_1,n_2)$ & $\delta$ & $d_2$ & $(0.25,0.25)$ & $(0.5,0.5)$ & $(0.25,0.5)$ &
$(0.25,0.25)$ & $(0.5,0.5)$ & $(0.25,0.5)$
  \\ \hline
$(10,10)$ & 1 & 4       & 0.002 & 0.002 & 0.002    &   0.000 & 0.002  &0.003  \\
$(10,10)$ & 0.75 & 2.25 & 0.005 & 0.005 & 0.009   & 0.007 & 0.011  & 0.016  \\
$(10,10)$ & 0.5 & 1     & 0.080 & 0.042 & 0.049   & 0.102 & 0.061  & 0.071 \\
\hline
$ (10,20)$ & 1  &  4    & 0.000 & 0.000 & 0.000   & 0.000 & 0.000  &0.000    \\
$ (10,20)$ & 0.75 & 2.25& 0.007 & 0.012 & 0.007   & 0.017 & 0.015  & 0.012  \\
$ (10,20)$ & 0.5 & 1    & 0.055 & 0.063 &  0.060  & 0.081 & 0.078  &0.084   \\
\hline
$(20,20)$ & 1 & 4       & 0.000 & 0.000 & 0.000   & 0.000 & 0.000  & 0.000  \\
$(20,20)$ & 0.75 & 2.25 & 0.000 & 0.001 & 0.002   & 0.017 & 0.003  & 0.006  \\
$(20,20)$ & 0.5 & 1     & 0.060 & 0.066 & 0.080  & 0.090 & 0.091  &0.096   \\
\hline
$(50,50)$ & 1 & 4       & 0.000 & 0.000 & 0.000    & 0.000 & 0.000  & 0.000  \\
$(50,50)$ & 0.75  & 2.25& 0.000 & 0.000 & 0.000   & 0.000  & 0.000  &0.001   \\
$(50,50)$ & 0.5 & 1     & 0.041 & 0.058 & 0.052    & 0.071 & 0.087  & 0.073 \\
\hline
\hline
\end{tabular}
\caption{\label{tab6} \it Simulated Type I error  rates of the asymptotic test \eqref{testl2asy} for the equivalence of two shifted EMAX models defined in \eqref{ex2}. The threshold in \eqref{hypothesesL2} is chosen as $\epsilon_2=1$}
}
\end{table}

%\FloatBarrier

In Tables \ref{tab7} and  \ref{tab8} we display the power of the two  tests under various alternatives specified by the value $
\beta_{1,1}=\delta$ in
model \eqref{ex2}.
We observe a reasonable power of both tests in all cases
under consideration.  In those  cases where the asymptotic test  \eqref{testl2asy} keeps (or exceeds) its nominal level it is slightly more
powerful than the bootstrap test \eqref{test}. The opposite performance can be observed in those cases where the
asymptotic test is conservative (e.g., if $\alpha=10\%,\ n_1=n_2=10$).  We also note that the power
of both tests is a decreasing function of the distance  $d_2$, as predicted by the asymptotic theory.

\begin{table}[!h]
  {\tiny
\centering
\begin{tabular}{||c|c|c|ccc||ccc||}
\hline
\multicolumn{1}{||c|}{} & \multicolumn{1}{c|}{} & \multicolumn{1}{|c|}{} & \multicolumn{3}{|c||}{$\alpha=0.05$} & \multicolumn{3}{|c||}{$\alpha=0.1$} \\\hline
\multicolumn{1}{||c|}{} & \multicolumn{1}{c|}{} & \multicolumn{1}{|c|}{} &
\multicolumn{3}{|c||}{$(\sigma_1^2,\sigma_2^2)$} & \multicolumn{3}{|c||}{$(\sigma_1^2,\sigma_2^2)$} \\\hline
$(n_1,n_2)$ & $\delta$ & $d_2$ & $(0.25,0.25)$ & $(0.5,0.5)$ & $(0.25,0.5)$ &
$(0.25,0.25)$ & $(0.5,0.5)$ & $(0.25,0.5)$
  \\ \hline
$(10,10)$ & 0.25 & 0.25  & 0.210 & 0.118 & 0.134  & 0.300 & 0.212 &0.256 \\
$(10,10)$ & 0.1 & 0.04   & 0.294 & 0.132 & 0.186   & 0.427 & 0.250 &0.312  \\
$(10,10)$ & 0 & 0        & 0.351 & 0.145 & 0.176   & 0.467 & 0.286 &0.340  \\
\hline
$ (10,20)$ & 0.25 & 0.25 & 0.257 & 0.125 &  0.191  & 0.392 & 0.234 & 0.305 \\
$ (10,20)$ & 0.1 & 0.04  & 0.395 & 0.164 &  0.254  & 0.535 & 0.305 &0.395 \\
$ (10,20)$ & 0 & 0       & 0.437 & 0.158 & 0.291  & 0.598 & 0.290 & 0.474  \\
\hline
$ (20,20)$ & 0.25 & 0.25 & 0.392 & 0.171 &  0.225  & 0.534 & 0.302 & 0.382 \\
$ (20,20)$ & 0.1 & 0.04  & 0.560 & 0.308 &  0.418   & 0.720 & 0.460 & 0.562 \\
$ (20,20)$ & 0 & 0       & 0.610 & 0.314 & 0.390   & 0.757 & 0.462 & 0.555  \\
\hline
$(50,50)$ & 0.25 & 0.25  & 0.724 & 0.460 & 0.554   & 0.825 & 0.595 &0.825  \\
$(50,50)$ & 0.1 & 0.04   & 0.961 & 0.691 & 0.821  & 0.982 & 0.824 & 0.973  \\
$(50,50)$ & 0 & 0        & 0.984 & 0.734 & 0.865  & 0.998 & 0.861 & 0.999  \\
\hline
\hline
\end{tabular}
\caption{\label{tab7} \it Simulated power of the bootstrap test \eqref{test} for the equivalence of two shifted EMAX models defined in \eqref{ex2}. The threshold in \eqref{hypothesesL2} is chosen as $\epsilon_2=1.$}
}
\end{table}

\begin{table}[!h]
  {\tiny
\centering
\begin{tabular}{||c|c|c|ccc||ccc||}
\hline
\multicolumn{1}{||c|}{} & \multicolumn{1}{c|}{} & \multicolumn{1}{|c|}{} & \multicolumn{3}{|c||}{$\alpha=0.05$} & \multicolumn{3}{|c||}{$\alpha=0.1$} \\\hline
\multicolumn{1}{||c|}{} & \multicolumn{1}{c|}{} & \multicolumn{1}{|c|}{} &
\multicolumn{3}{|c||}{$(\sigma_1^2,\sigma_2^2)$} & \multicolumn{3}{|c||}{$(\sigma_1^2,\sigma_2^2)$} \\\hline
$(n_1,n_2)$ & $\delta$ & $d_2$ & $(0.25,0.25)$ & $(0.5,0.5)$ & $(0.25,0.5)$ &
$(0.25,0.25)$ & $(0.5,0.5)$ & $(0.25,0.5)$
  \\ \hline
$(10,10)$ & 0.25 & 0.25  & 0.264 & 0.103 & 0.175 & 0.311 & 0.131 &0.217   \\
$(10,10)$ & 0.1 & 0.04   & 0.351 & 0.139 & 0.196   & 0.431 & 0.183 &0.247 \\
$(10,10)$ & 0 & 0        & 0.381 & 0.120 & 0.222  & 0.468 & 0.168 &0.279  \\
\hline
$ (10,20)$ & 0.25 & 0.25 & 0.305 & 0.147 &  0.256 & 0.382 & 0.192 &0.317  \\
$ (10,20)$ & 0.1 & 0.04  & 0.468 & 0.218 &  0.359 & 0.536 & 0.268 &0.438  \\
$ (10,20)$ & 0 & 0       & 0.510 & 0.220 & 0.358  & 0.570 &0.272  & 0.455  \\
\hline
$(20,20)$ & 0.25 & 0.25  & 0.423 & 0.271 & 0.321  & 0.493 & 0.328 &0.341   \\
$(20,20)$ & 0.1 & 0.04   & 0.640 & 0.328 & 0.501   & 0.716 & 0.407 & 0.585 \\
$(20,20)$ & 0 & 0        & 0.690 & 0.351 & 0.501  & 0.781 & 0.438 &0.573  \\
\hline
$(50,50)$ & 0.25 & 0.25  & 0.659 & 0.475 & 0.534  & 0.740 & 0.562 &0.649  \\
$(50,50)$ & 0.1 & 0.04   & 0.965 & 0.750 & 0.868  & 0.974 & 0.813 & 0.911  \\
$(50,50)$ & 0 & 0        & 0.980 & 0.848 & 0.937  & 0.991 & 0.893 & 0.946  \\
\hline\hline
\end{tabular}
\caption{\label{tab8} \it Simulated power of the asymptotic test \eqref{testl2asy} for the equivalence of two shifted EMAX models defined in \eqref{ex2}. The threshold in \eqref{hypothesesL2} is chosen as $\epsilon_2=1$.}
}
\end{table}
%\FloatBarrier

\subsection{Tests based on the distance $d_\infty$} \label{sec52}

We now investigate the maximum deviation distance and also provide a comparison with the test proposed by \cite{gsteiger2011}.
Motivated by  the discussion in Section \ref{sec4}  we distinguish the cases where the cardinality of the set $\mathcal{E}$ is one or larger than one.
The results will show that with an increasing size of the set $\mathcal{E} $
 the test is getting more conservative.

\begin{example} \label{exam1} {\rm
{\bf {($\#\mathcal{E} =1$)}}  We begin with a comparison of an EMAX with an exponential model, that is
\begin{equation}\label{ex3a}
m_1(x,\beta_1)=\beta_{11} +\frac{\beta_{12}x}{\beta_{13}+x}  \qquad \mbox{and} \qquad m_2(x,\beta_2)=  \beta_{21}+ \beta_{22}\cdot(\exp{(\tfrac{x}{ \beta_{23}})}-1),
\end{equation}
where $\beta_{1} =  ( \beta_{11}, \beta_{12}, \beta_{13})  =(1,2 ,1)$ and $\beta_{2} =  (  \beta_{21}, \beta_{22}, \beta_{23} ) =(\delta , 2.2,8)$
[see Figure \ref{fig4}].
\begin{figure}[h]
	\centering
		\includegraphics[width=0.4\textwidth]{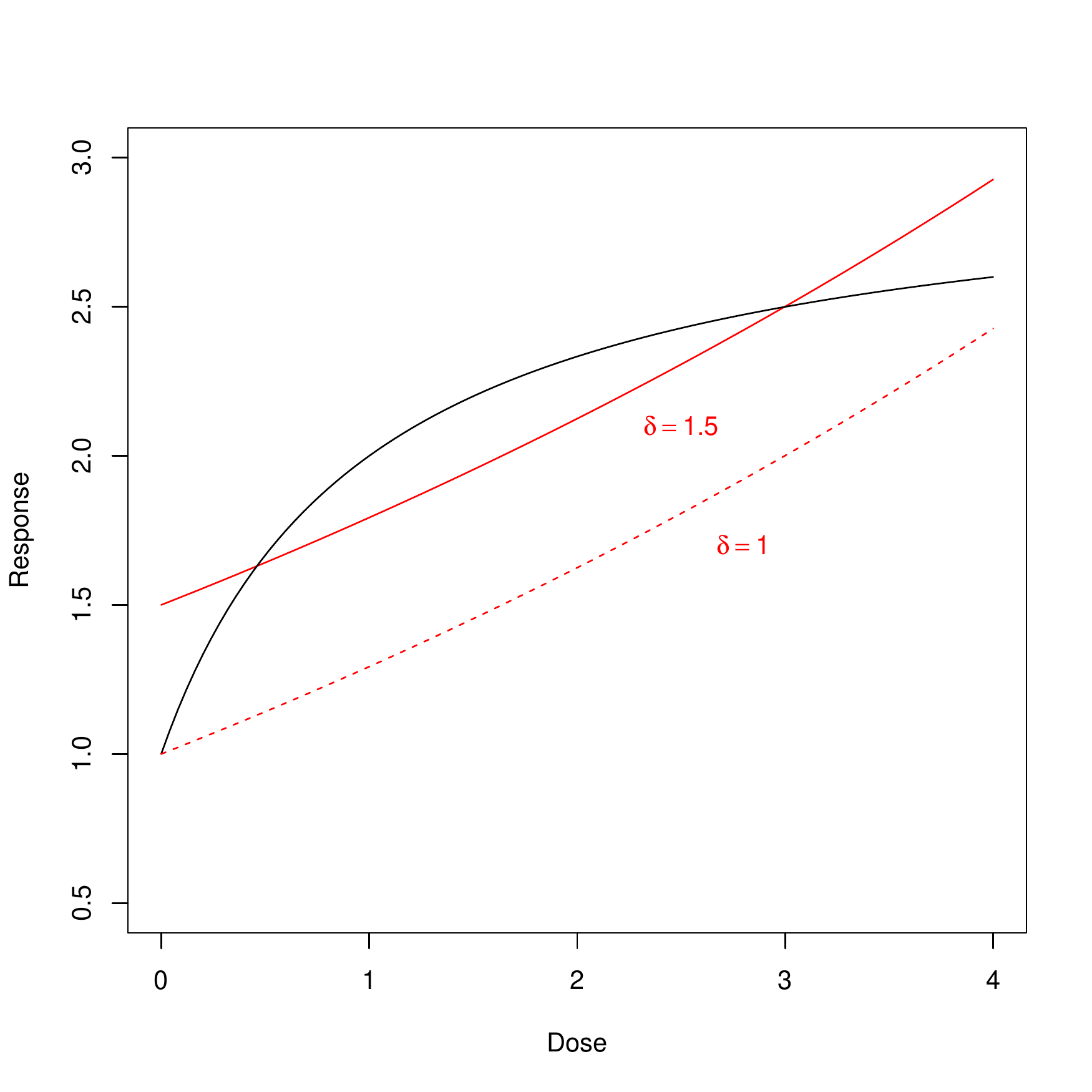}
	\caption{\it The models defined in \eqref{ex3a} for two different choices of $\delta$.}	
	\label{fig4}
\end{figure}
In Table  \ref{tab11a} we display the simulated rejection probabilities of
the bootstrap test  \eqref{testinf} under  the  null hypothesis in \eqref{h0inf}, where  $\epsilon_\infty=1$.  \\
If the cardinality of the set $\mathcal{E}$ is one, the distribution of the test statistic $\hat d_\infty$ is a centered normal distribution
with variance defined in \eqref{unique}.  Thus, if the unique extremal point has been estimated, we obtain
an estimate, say $\hat \sigma_\infty^2,$  of
the  asymptotic variance of the statistic $\hat d_\infty$. The null hypothesis
is now rejected (at asymptotic level $\alpha$), whenever
\begin{equation} \label{testunique}
\hat d_\infty <  \varepsilon_\infty +  { \hat \sigma_\infty  \over \sqrt n} u_\alpha~,
\end{equation}
where $u_\alpha$ is the $\alpha-$quantile of the standard normal distribution.
The results for this  test are given in  Table \ref{tab13a}.
We observe that the bootstrap test \eqref{testinf} keeps its nominal level at the boundary of the null hypothesis, whereas
the level is smaller in the interior (this confirms the theoretical results from Section \ref{sec4}).  The approximation is less precise for small sample sizes. Compared to the bootstrap test based on the distance $d_2$ the test \eqref{testinf} is conservative.
The asymptotic test \eqref{testunique}
is very conservative, even for relative large sample sizes (see Table \ref{tab13a}).
A possible explanation for this observation consists in the fact that the estimation of the extremal point is a difficult problem. 
In Table  \ref{tab11a} we also display the rejection probabilities of the
test of \cite{gsteiger2011}  in brackets. This test is very conservative as its level is practically $0$
for almost all cases under consideration.   \\
The simulated  power of the bootstrap and the asymptotic $d_\infty$-test is  displayed in Table \ref{tab12a} and \ref{tab14a}.  We observe a substantially better performance
of the bootstrap test  \eqref{testinf} in all cases of consideration. In Table \ref{tab12a} we also display the rejection probabilities of the
 test of \cite{gsteiger2011}  in brackets and we conclude that the methods proposed in this paper yield a substantial improvement for small sample sizes or large variances.
 }
\end{example}

\begin{table}[!h]
  {\tiny
\centering
\begin{tabular}{||c|c|c|ccc||ccc||}
\hline
\multicolumn{1}{||c|}{} & \multicolumn{1}{c|}{} & \multicolumn{1}{|c|}{} & \multicolumn{3}{|c||}{$\alpha=0.05$} & \multicolumn{3}{|c||}{$\alpha=0.1$} \\\hline
\multicolumn{1}{||c|}{} & \multicolumn{1}{c|}{} & \multicolumn{1}{|c|}{} &
\multicolumn{3}{|c||}{$(\sigma_1^2,\sigma_2^2)$} & \multicolumn{3}{|c||}{$(\sigma_1^2,\sigma_2^2)$} \\\hline
$(n_1,n_2)$ & $\delta$ & $d_\infty$ & $(0.25,0.25)$ & $(0.5,0.5)$ & $(0.25,0.5)$ &
$(0.25,0.25)$ & $(0.5,0.5)$ & $(0.25,0.5)$
  \\ \hline
$(10,10)$ & 0.25 & 1.5 & 0.001 (0.000)& 0.001 (0.000)&  0.000 (0.000) & 0.000 (0.000)& 0.004 (0.000)&  0.000 (0.000) \\
$(10,10)$ & 0.5 & 1.25 & 0.005 (0.000)& 0.011 (0.000)&  0.006 (0.000) & 0.013 (0.000)& 0.030 (0.000)&  0.020 (0.000) \\
$(10,10)$ & 0.75& 1    & 0.045 (0.007)& 0.037 (0.000)&  0.036 (0.001) & 0.102 (0.021)& 0.086 (0.002)&  0.090 (0.007) \\
\hline
$(10,20)$ & 0.25 & 1.5 & 0.000 (0.000)& 0.002 (0.000)&  0.000 (0.000) & 0.000 (0.000)& 0.002 (0.000)&  0.000 (0.000) \\
$(10,20)$ & 0.5 & 1.25 & 0.004 (0.000)& 0.013 (0.000)&  0.005 (0.000) & 0.015 (0.000)& 0.025 (0.000)&  0.009 (0.000)  \\
$(10,20)$ & 0.75 &1    & 0.045 (0.017)& 0.046 (0.002)&  0.028 (0.004) & 0.099 (0.042)& 0.104 (0.011)&  0.079 (0.017)  \\
\hline
$(20,20)$ & 0.25 & 1.5 & 0.000 (0.000)& 0.000 (0.000)&  0.000 (0.000) & 0.000 (0.000)& 0.000 (0.000)& 0.000 (0.000) \\
$(20,20)$ & 0.5 & 1.25 & 0.001 (0.000)& 0.004 (0.000)&  0.000 (0.000) & 0.006 (0.000)& 0.018 (0.000)& 0.011 (0.000) \\
$(20,20)$ & 0.75 &1    & 0.034 (0.013)& 0.038 (0.003)&  0.048 (0.007) & 0.091 (0.033)& 0.100 (0.020)& 0.104 (0.026) \\
\hline
$(50,50)$ & 0.25 & 1.5 & 0.000 (0.000)& 0.000 (0.000)&  0.000 (0.000)& 0.000 (0.000)& 0.000 (0.000)& 0.000 (0.000)  \\
$(50,50)$ & 0.5 & 1.25 & 0.000 (0.000)& 0.000 (0.000)&  0.000 (0.000) & 0.002 (0.000)& 0.001 (0.000)& 0.000 (0.000)  \\
$(50,50)$ & 0.75 &1    & 0.051 (0.013)& 0.059 (0.009)&  0.058 (0.010) & 0.096 (0.040)& 0.108 (0.043)& 0.105 (0.096)  \\
\hline
\hline
\end{tabular}
\caption{\label{tab11a} \it Simulated Type I error rates  of the bootstrap test  \eqref{testinf}  for the equivalence of an
EMAX and an exponential model defined by \eqref{ex3a}. The threshold in \eqref{h0inf} is chosen as $\epsilon_\infty=1$. The numbers in brackets show the simulated
Type I error  rates of the test proposed in \cite{gsteiger2011}. }
}
\end{table}

\begin{table}[!h]
  {\tiny
\centering
\begin{tabular}{||c|c|c|ccc||ccc||}
\hline
\multicolumn{1}{||c|}{} & \multicolumn{1}{c|}{} & \multicolumn{1}{|c|}{} & \multicolumn{3}{|c||}{$\alpha=0.05$} & \multicolumn{3}{|c||}{$\alpha=0.1$} \\\hline
\multicolumn{1}{||c|}{} & \multicolumn{1}{c|}{} & \multicolumn{1}{|c|}{} &
\multicolumn{3}{|c||}{$(\sigma_1^2,\sigma_2^2)$} & \multicolumn{3}{|c||}{$(\sigma_1^2,\sigma_2^2)$} \\\hline
$(n_1,n_2)$ & $\delta$ & $d_\infty$ & $(0.25,0.25)$ & $(0.5,0.5)$ & $(0.25,0.5)$ &
$(0.25,0.25)$ & $(0.5,0.5)$ & $(0.25,0.5)$
  \\ \hline
$(10,10)$ & 0.25 & 1.5 & 0.000 & 0.000 &  0.000 & 0.000 & 0.000 &  0.000  \\
$(10,10)$ & 0.5 & 1.25 & 0.001 & 0.001 &  0.000 & 0.003 & 0.004 &  0.001 \\
$(10,10)$ & 0.75& 1    & 0.012 & 0.005 &  0.003  & 0.029 & 0.010 &  0.001 \\
\hline
$(10,20)$ & 0.25 & 1.5 & 0.000 & 0.000 &  0.000  & 0.000 & 0.000 &  0.000\\
$(10,20)$ & 0.5 & 1.25 & 0.000 & 0.005 &  0.001 & 0.000 & 0.007 &  0.003 \\
$(10,20)$ & 0.75 &1    & 0.019 & 0.006 &  0.009  & 0.038 & 0.014 &  0.023 \\
\hline
$(20,20)$ & 0.25 & 1.5 & 0.000 & 0.000 &  0.000 & 0.000 & 0.000 &  0.000 \\
$(20,20)$ & 0.5 & 1.25 & 0.000 & 0.000 &  0.001 & 0.000 & 0.001 &  0.001 \\
$(20,20)$ & 0.75 &1    & 0.011 & 0.036 & 0.009  & 0.033 & 0.025 &  0.027   \\
\hline
$(50,50)$ & 0.25 & 1.5 & 0.000 & 0.000 &  0.000 & 0.000 & 0.000 & 0.000 \\
$(50,50)$ & 0.5 & 1.25 & 0.000 & 0.000 &  0.000 & 0.000 & 0.003 & 0.000 \\
$(50,50)$ & 0.75 &1    & 0.016 & 0.015 &  0.012  & 0.039 & 0.039 & 0.041 \\
\hline
\hline
\end{tabular}
\caption{\label{tab13a} \it Simulated Type I error  rates of the asymptotic test \eqref{testunique} for the equivalence of
an EMAX and an exponential model defined by \eqref{ex3a}. The threshold in \eqref{h0inf} is chosen as $\epsilon_\infty=1$}
}
\end{table}

\begin{table}[!h]
  {\tiny
\centering
\begin{tabular}{||c|c|c|ccc||ccc||}
\hline
\multicolumn{1}{||c|}{} & \multicolumn{1}{c|}{} & \multicolumn{1}{|c|}{} & \multicolumn{3}{|c||}{$\alpha=0.05$} & \multicolumn{3}{|c||}{$\alpha=0.1$} \\\hline
\multicolumn{1}{||c|}{} & \multicolumn{1}{c|}{} & \multicolumn{1}{|c|}{} &
\multicolumn{3}{|c||}{$(\sigma_1^2,\sigma_2^2)$} & \multicolumn{3}{|c||}{$(\sigma_1^2,\sigma_2^2)$} \\\hline
$(n_1,n_2)$ & $\delta$ & $d_\infty$ & $(0.25,0.25)$ & $(0.5,0.5)$ & $(0.25,0.5)$ &
$(0.25,0.25)$ & $(0.5,0.5)$ & $(0.25,0.5)$
  \\ \hline
$(10,10)$ & 1 & 0.75  & 0.160 (0.026) & 0.093 (0.004)&  0.125 (0.007) & 0.297 (0.083)& 0.225 (0.007)&  0.229 (0.033) \\
$(10,10)$ & 1.5 & 0.5 & 0.237 (0.037) & 0.133 (0.003)&  0.164 (0.009) & 0.383 (0.117)& 0.231 (0.018)&  0.309 (0.029) \\
%$(10,10)$ &   & 0.25 &   &  &    & & &    \\
\hline
$(10,20)$ & 1 &0.75   & 0.185 (0.084) & 0.123 (0.006)&  0.159 (0.025)  & 0.320 (0.162)& 0.226 (0.035)&  0.283 (0.089) \\
$(10,20)$ & 1.5 &0.5  & 0.300 (0.087) & 0.175 (0.005)&  0.269 (0.035) & 0.457 (0.190)& 0.305 (0.043)&  0.414 (0.120) \\
%$(10,20)$ &   & 0.25 &   &  &    & & &    \\
\hline
$(20,20)$ & 1 &0.75   & 0.214 (0.130) & 0.138 (0.022)&  0.171 (0.054) & 0.393 (0.248)& 0.271 (0.086)&  0.345 (0.137) \\
$(20,20)$ & 1.5 & 0.5 & 0.401 (0.190) & 0.229 (0.036)&  0.363 (0.080) & 0.604 (0.356)& 0.398 (0.122)&  0.523 (0.189) \\
% $(20,20)$ &   & 0.25 &   &  &    & & &    \\
\hline
$(50,50)$ & 1 &0.75   & 0.504 (0.400) & 0.274 (0.183)&  0.363 (0.297)& 0.662 (0.552)& 0.416 (0.326)&  0.532 (0.433) \\
$(50,50)$ & 1.5 &0.5  & 0.777 (0.667) & 0.491 (0.294)&  0.606 (0.493) & 0.877 (0.791)& 0.648 (0.478)&  0.739 (0.604) \\
%$(50,50)$ &   & 0.25 &   &  &    & & &    \\
\hline
\hline
\end{tabular}
\caption{\label{tab12a} \it Simulated power of the bootstrap test  \eqref{testinf}
 for the equivalence of an EMAX and an exponential model defined by \eqref{ex3a}. The threshold in \eqref{h0inf} is chosen as $\epsilon_\infty=1$. The numbers in brackets show the simulated power of the test proposed in \cite{gsteiger2011}.  }
}
\end{table}
\begin{table}[!h]
  {\tiny
\centering
\begin{tabular}{||c|c|c|ccc||ccc||}
\hline
\multicolumn{1}{||c|}{} & \multicolumn{1}{c|}{} & \multicolumn{1}{|c|}{} & \multicolumn{3}{|c||}{$\alpha=0.05$} & \multicolumn{3}{|c||}{$\alpha=0.1$} \\\hline
\multicolumn{1}{||c|}{} & \multicolumn{1}{c|}{} & \multicolumn{1}{|c|}{} &
\multicolumn{3}{|c||}{$(\sigma_1^2,\sigma_2^2)$} & \multicolumn{3}{|c||}{$(\sigma_1^2,\sigma_2^2)$} \\\hline
$(n_1,n_2)$ & $\delta$ & $d_\infty$ & $(0.25,0.25)$ & $(0.5,0.5)$ & $(0.25,0.5)$ &
$(0.25,0.25)$ & $(0.5,0.5)$ & $(0.25,0.5)$
  \\ \hline
$(10,10)$ & 1 & 0.75  & 0.042 & 0.006 &  0.011  & 0.109 & 0.017 &  0.046\\
$(10,10)$ & 1.5 & 0.5 & 0.064 & 0.008 &  0.014 & 0.140 & 0.026 &  0.047 \\
%$(10,10)$ &   & 0.25 &   &  &    & & &    \\
\hline
$(10,20)$ & 1 &0.75   & 0.114 & 0.014 & 0.048  & 0.199 & 0.047 &  0.106 \\
$(10,20)$ & 1.5 &0.5  & 0.129 & 0.018 & 0.059  & 0.228 & 0.052 &  0.127 \\
%$(10,20)$ &   & 0.25 &   &  &    & & &    \\
\hline
$(20,20)$ & 1 &0.75   & 0.151 & 0.036 &  0.064 & 0.285 & 0.093 &  0.170 \\
$(20,20)$ & 1.5 & 0.5 & 0.209 & 0.060 &  0.104  & 0.360 & 0.120 &  0.202 \\
%$(20,20)$ &   & 0.25 &   &  &    & & &    \\
\hline
$(50,50)$ & 1 &0.75   & 0.417 & 0.206 &  0.303 & 0.569 & 0.337 & 0.440 \\
$(50,50)$ & 1.5 &0.5  & 0.706 & 0.267 &  0.408 & 0.826 & 0.462 & 0.630 \\
%$(50,50)$ &   & 0.25 &   &  &    & & &    \\
\hline
\hline
\end{tabular}
\caption{\label{tab14a} \it Simulated power of the asymptotic test  \eqref{testunique}
 for the equivalence of an EMAX and an exponential model defined by \eqref{ex3a}. The threshold in \eqref{h0inf} is chosen as $\epsilon_\infty=1$.}
}
\end{table}
% \subsubsection{Finite sample properties for arbitrary $\mathcal{E}$}
In the remaining part of this section we consider three further scenarios, where the true maximum absolute distance of the models
$m_1$ and $m_2$ is attained at more than one point.
 In this case,
an asymptotic test based on the maximum deviation is not available and therefore only the bootstrap test can be used.
Our results demonstrate that the  test is  conservative compared to the scenario before, where
$\#\mathcal{E}=1$.
This  confirms our theoretical findings in Theorem \ref{thm43}. Moreover, the test gets more conservative if the size of the set
$\mathcal{E}$ increases. We also display (the numbers in brackets) the corresponding values for the test of \cite{gsteiger2011},
which is very conservative and less powerful than the bootstrap test in all cases under consideration.

\begin{example} {\bf  ($\# \mathcal{E} =2 $)} \label{exam2} {\rm
We consider two EMAX models, given by
\begin{equation}
m_1(x,\beta_1)=\beta_{11} +\frac{\beta_{12} x}{\beta_{13}+x}\text{ and } m_2(x,\beta_2)=\beta_{21}+\frac{\beta_{22} x}{\beta_{23}+x},\label{ex3}
\end{equation}
where $\beta_{1} =  ( \beta_{11}, \beta_{12}, \beta_{13})  =(\delta ,6, 2)$ and $\beta_{2} =  (  \beta_{21}, \beta_{22}, \beta_{23} ) =(0,5,1)$.
In this case  the  maximum absolute difference of $d_\infty=\delta$ is attained at  the boundary points of the design space,
that is $\mathcal{E}=\{0,4\}$.
The corresponding  rejection probabilities under the null hypothesis are presented in
Table \ref{tab11}. We observe that the bootstrap test keeps  its level in all situations, but it is conservative
(see also Theorem \ref{thm43}). The test of \cite{gsteiger2011} is even more conservative.
We also observe an improvement in power by the new test in comparison with the test
of \cite{gsteiger2011}, in particular for small samples sizes (see Table \ref{tab12}).
}
\end{example}

\begin{table}[!h]
  {\tiny
\centering
\begin{tabular}{||c|c|ccc||ccc||}
\hline
\multicolumn{1}{||c|}{} &   \multicolumn{1}{|c|}{} & \multicolumn{3}{|c||}{$\alpha=0.05$} & \multicolumn{3}{|c||}{$\alpha=0.1$} \\\hline
\multicolumn{1}{||c|}{}   & \multicolumn{1}{|c|}{} &
\multicolumn{3}{|c||}{$(\sigma_1^2,\sigma_2^2)$} & \multicolumn{3}{|c||}{$(\sigma_1^2,\sigma_2^2)$} \\\hline
$(n_1,n_2)$   & $d_\infty=\delta$ & $(0.25,0.25)$ & $(0.5,0.5)$ & $(0.25,0.5)$ &
$(0.25,0.25)$ & $(0.5,0.5)$ & $(0.25,0.5)$
  \\ \hline
$(10,10)$ & 2    & 0.000 (0.000) & 0.000 (0.000)& 0.000 (0.000)& 0.000 (0.000)& 0.001 (0.000)& 0.001 (0.000)\\
$(10,10)$ & 1.5 & 0.000 (0.000)& 0.000 (0.000)& 0.000 (0.000)& 0.000 (0.000)& 0.004 (0.000)& 0.002 (0.000)\\
$(10,10)$ & 1  & 0.018 (0.002)& 0.017 (0.001)& 0.008 (0.000)& 0.042 (0.006)& 0.042 (0.003)& 0.034 (0.000)\\
\hline
$ (10,20)$ & 2   &  0.000 (0.000)& 0.000 (0.000)& 0.000 (0.000)& 0.000 (0.000)& 0.000 (0.000)& 0.000 (0.000)\\
$ (10,20)$ & 1.5 &  0.000 (0.000)& 0.000 (0.000)& 0.000 (0.000)& 0.001 (0.000)& 0.000 (0.000)& 0.003 (0.000)\\
$ (10,20)$ & 1 &  0.012 (0.002)& 0.015 (0.000)& 0.009 (0.000)& 0.052 (0.005)& 0.049 (0.003)& 0.042 (0.000)\\
\hline
$(20,20)$ & 2    & 0.000 (0.000)& 0.000 (0.000)& 0.000 (0.000)& 0.000 (0.000)& 0.002 (0.000)& 0.000 (0.000)\\
$(20,20)$ & 1.5 & 0.000 (0.000)& 0.000 (0.000)& 0.000 (0.000)& 0.001 (0.000)& 0.001 (0.000)& 0.000 (0.000)\\
$(20,20)$ & 1  & 0.023 (0.000)& 0.012 (0.000)& 0.015 (0.003)& 0.066 (0.008)& 0.043 (0.001)& 0.042 (0.004)\\
\hline
$(50,50)$ & 2    &  0.000 (0.000)& 0.000 (0.000)& 0.000 (0.000)& 0.000 (0.000)& 0.001 (0.000)& 0.000 (0.000)\\
$(50,50)$ & 1.5 &  0.000 (0.000)& 0.000 (0.000)& 0.000 (0.000)& 0.000 (0.000)& 0.001 (0.000)& 0.000 (0.000)\\
$(50,50)$ & 1  &  0.022 (0.002)& 0.020 (0.000)& 0.016 (0.002)& 0.050 (0.007)& 0.048 (0.006)& 0.049 (0.007)\\
\hline
\hline
\end{tabular}
\caption{\label{tab11} \it Simulated Type I error  rates of the bootstrap $d_\infty$-test
 \eqref{testinf}  for the equivalence of two EMAX-models defined by \eqref{ex3}.  The threshold in \eqref{h0inf} is chosen as $\epsilon_\infty=1$. The numbers in brackets show the simulated Type I error  rates of the test proposed by \cite{gsteiger2011}.}
}
\end{table}

 \begin{table}[!h]
  {\tiny
\centering
\begin{tabular}{||c|c|ccc||ccc||}
\hline
\multicolumn{1}{||c|}{} &   \multicolumn{1}{|c|}{} & \multicolumn{3}{|c||}{$\alpha=0.05$} & \multicolumn{3}{|c||}{$\alpha=0.1$} \\\hline
\multicolumn{1}{||c|}{}   & \multicolumn{1}{|c|}{} &
\multicolumn{3}{|c||}{$(\sigma_1^2,\sigma_2^2)$} & \multicolumn{3}{|c||}{$(\sigma_1^2,\sigma_2^2)$} \\\hline
$(n_1,n_2)$   & $d_\infty=\delta$ & $(0.25,0.25)$ & $(0.5,0.5)$ & $(0.25,0.5)$ &
$(0.25,0.25)$ & $(0.5,0.5)$ & $(0.25,0.5)$
  \\ \hline
$(10,10)$ & 0.75  & 0.074 (0.011)& 0.042 (0.003)& 0.059 (0.002)& 0.154 (0.040)& 0.116 (0.007)& 0.133 (0.021)\\
$(10,10)$ & 0.5   & 0.189 (0.055)& 0.111 (0.001)& 0.139 (0.011)& 0.312 (0.137)& 0.202 (0.014)& 0.249 (0.049)\\
$(10,10)$ & 0     & 0.267 (0.088)& 0.116 (0.003)& 0.170 (0.020)& 0.415 (0.211)& 0.237 (0.026)& 0.299 (0.070)\\
\hline
$ (10,20)$ & 0.75 & 0.067 (0.015)& 0.053 (0.002)& 0.070 (0.007)& 0.161 (0.045)& 0.124 (0.019)& 0.156 (0.030)\\
$ (10,20)$ & 0.5  & 0.229 (0.106)& 0.141 (0.006)& 0.185 (0.057)& 0.373 (0.231)& 0.249 (0.036)& 0.320 (0.138)\\
$ (10,20)$ & 0    & 0.343 (0.172)& 0.191 (0.011)& 0.234 (0.062)& 0.513 (0.314)& 0.314 (0.055)& 0.376 (0.168)\\
\hline
$(20,20)$ & 0.75  & 0.120 (0.045)& 0.050 (0.005)& 0.074 (0.023)& 0.215 (0.102)& 0.149 (0.028)& 0.184 (0.068)\\
$(20,20)$ & 0.5   & 0.372 (0.239)& 0.192 (0.032)& 0.243 (0.079)& 0.531 (0.380)& 0.324 (0.117)& 0.367 (0.192)\\
$(20,20)$ & 0     & 0.462 (0.334)& 0.234 (0.049)& 0.338 (0.113)& 0.591 (0.511)& 0.392 (0.148)& 0.487 (0.260)\\
\hline
$(50,50)$ & 0.75  & 0.231 (0.133)& 0.110 (0.040)& 0.154 (0.069)& 0.368 (0.247)& 0.219 (0.108)& 0.296 (0.170)\\
$(50,50)$ & 0.5   & 0.708 (0.613)& 0.387 (0.294)& 0.542 (0.411)& 0.834 (0.770)& 0.554 (0.469)& 0.689 (0.593)\\
$(50,50)$ & 0     & 0.792 (0.773)& 0.528 (0.468)& 0.630 (0.576)& 0.873 (0.862)& 0.669 (0.627)& 0.757 (0.721)\\
\hline\hline
\end{tabular}
\caption{\label{tab12} \it Simulated power of the bootstrap $d_\infty$-test
 \eqref{testinf}  for the equivalence of two EMAX-models defined by \eqref{ex3}. The threshold in \eqref{h0inf} is chosen as $\epsilon_\infty=1$. The numbers in brackets show the simulated power of the test proposed in \cite{gsteiger2011}.}
}
\end{table}

\begin{example} {\bf  ($\# \mathcal{E} =3 $)} \label{exam3} {\rm
In the next scenario we consider  a quadratic and a linear model, that is\be
m_1(x,\beta_1)=  \beta_{11} x^2 +  \beta_{12} x+ \beta_{13} \text{ and }  m_2(x,\beta_2)= \beta_{21} x+  \beta_{22}. \label{ex4}
\ee
where $\beta_{1} =  ( \beta_{11}, \beta_{12}, \beta_{13})  =(\delta,-3 \delta ,3 \delta)$ and $\beta_{2} =  (  \beta_{21}, \beta_{22}) =(\delta , \delta)$.
The maximum absolute distance is given by $d_\infty=2\delta$, attained at $\mathcal{E}=\{0,2,4\}$.
The simulated rejection probabilities under the null hypothesis are shown in Table \ref{tab13}.
A comparison with Table \ref{tab12} shows  that the level decreases with the size of the  set of extremal points $\mathcal{E}$.
The results displayed in
Table \ref{tab14}  show again  that the new test has a larger power than  the test  of \cite{gsteiger2011}.
}
\end{example}

\begin{table}[!h]
  {\tiny
\centering
\begin{tabular}{||c|c|ccc||ccc||}
\hline
\multicolumn{1}{||c|}{} &   \multicolumn{1}{|c|}{} & \multicolumn{3}{|c||}{$\alpha=0.05$} & \multicolumn{3}{|c||}{$\alpha=0.1$} \\\hline
\multicolumn{1}{||c|}{}   & \multicolumn{1}{|c|}{} &
\multicolumn{3}{|c||}{$(\sigma_1^2,\sigma_2^2)$} & \multicolumn{3}{|c||}{$(\sigma_1^2,\sigma_2^2)$} \\\hline
$(n_1,n_2)$   & $d_\infty=2\delta$ & $(0.25,0.25)$ & $(0.5,0.5)$ & $(0.25,0.5)$ &
$(0.25,0.25)$ & $(0.5,0.5)$ & $(0.25,0.5)$
  \\ \hline
$(10,10)$ & 2   & 0.000 (0.000)& 0.001 (0.000)& 0.000 (0.000)& 0.006 (0.000)& 0.010 (0.000)& 0.023 (0.000)\\
$(10,10)$ & 1.5 & 0.000 (0.000)& 0.001 (0.000)& 0.005 (0.000)& 0.009 (0.000)& 0.009 (0.000)& 0.024 (0.000)\\
$(10,10)$ & 1   & 0.007 (0.000)& 0.006 (0.000)& 0.009 (0.000)& 0.028 (0.002)& 0.025 (0.000)& 0.038 (0.001)\\
\hline
$ (10,20)$ & 2   &  0.000 (0.000)& 0.000 (0.000)& 0.000 (0.000)& 0.001 (0.000)& 0.001 (0.000)& 0.006 (0.000)\\
$ (10,20)$ & 1.5 &  0.000 (0.000)& 0.000 (0.000)& 0.000 (0.000)& 0.002 (0.000)& 0.002 (0.000)& 0.006 (0.000)\\
$ (10,20)$ & 1 &  0.008 (0.000)& 0.009 (0.000)& 0.006 (0.000)& 0.028 (0.004)& 0.023 (0.000)& 0.022 (0.004)\\
\hline
$(20,20)$ & 2   & 0.000 (0.000)& 0.000 (0.000)& 0.004 (0.000)& 0.006 (0.000)& 0.006 (0.000)& 0.024 (0.000)\\
$(20,20)$ & 1.5 & 0.000 (0.000)& 0.000 (0.000)& 0.001 (0.000)& 0.006 (0.000)& 0.001 (0.000)& 0.018 (0.000)\\
$(20,20)$ & 1   & 0.005 (0.000)& 0.009 (0.000)& 0.003 (0.000)& 0.015 (0.001)& 0.030 (0.000)& 0.027 (0.000)\\
\hline
$(50,50)$ & 2    & 0.000 (0.000)& 0.000 (0.000)& 0.001 (0.000)& 0.004 (0.000)& 0.001 (0.000)& 0.015 (0.000)\\
$(50,50)$ & 1.5 & 0.000 (0.000)& 0.000 (0.000)& 0.001 (0.000)& 0.005 (0.000)& 0.006 (0.000)& 0.007 (0.000)\\
$(50,50)$ & 1  & 0.001 (0.000)& 0.004 (0.000)& 0.002 (0.000)& 0.014 (0.001)& 0.019 (0.000)& 0.027 (0.001)\\
\hline
\hline
\end{tabular}
\caption{\label{tab13} \it Simulated Type I error  rates of the bootstrap $d_\infty$-test
 \eqref{testinf}  for the equivalence of a quadratic and a linear model defined by \eqref{ex4}.  The threshold in \eqref{h0inf} is chosen as $\epsilon_\infty=1$. The numbers in brackets show the simulated Type I error  rates of the test proposed by \cite{gsteiger2011}.}
}
\end{table}

 \begin{table}[!h]
  {\tiny
\centering
\begin{tabular}{||c|c|ccc||ccc||}
\hline
\multicolumn{1}{||c|}{} &   \multicolumn{1}{|c|}{} & \multicolumn{3}{|c||}{$\alpha=0.05$} & \multicolumn{3}{|c||}{$\alpha=0.1$} \\\hline
\multicolumn{1}{||c|}{}   & \multicolumn{1}{|c|}{} &
\multicolumn{3}{|c||}{$(\sigma_1^2,\sigma_2^2)$} & \multicolumn{3}{|c||}{$(\sigma_1^2,\sigma_2^2)$} \\\hline
$(n_1,n_2)$   & $d_\infty=2\delta$ & $(0.25,0.25)$ & $(0.5,0.5)$ & $(0.25,0.5)$ &
$(0.25,0.25)$ & $(0.5,0.5)$ & $(0.25,0.5)$
  \\ \hline
$(10,10)$ & 0.4  & 0.290 (0.106)& 0.143 (0.011)& 0.213 (0.038)& 0.452 (0.210)& 0.256 (0.010)& 0.356 (0.105)\\
$(10,10)$ & 0.2  & 0.438 (0.198)& 0.206 (0.014)& 0.324 (0.088)& 0.619 (0.371)& 0.367 (0.083)& 0.473 (0.221)\\
\hline
$ (10,20)$ & 0.4 & 0.313 (0.158)& 0.176 (0.023)& 0.284 (0.100)& 0.484 (0.314)& 0.295 (0.089)& 0.446 (0.202)\\
$ (10,20)$ & 0.2  & 0.542 (0.329)& 0.262 (0.049)& 0.442 (0.197)& 0.694 (0.531)& 0.423 (0.140)& 0.602 (0.397)\\
\hline
$(20,20)$ & 0.4  & 0.500 (0.356)& 0.230 (0.092)& 0.377 (0.162)& 0.641 (0.572)& 0.377 (0.228)& 0.531 (0.344)\\
$(20,20)$ & 0.2  & 0.748 (0.661)& 0.430 (0.212)& 0.620 (0.415)& 0.858 (0.806)& 0.601 (0.406)& 0.764 (0.627)\\
\hline
$(50,50)$ & 0.4  & 0.879 (0.851)& 0.573 (0.448)& 0.767 (0.690)& 0.942 (0.926)& 0.733 (0.634)& 0.877 (0.821)\\
$(50,50)$ & 0.2   & 0.991 (0.986)& 0.828 (0.800)& 0.936 (0.929)& 0.998 (0.995)& 0.915 (0.899)& 0.973 (0.975)\\
\hline\hline
\end{tabular}
\caption{\label{tab14} \it Simulated power of the bootstrap $d_\infty$-test
 \eqref{testinf}  for the equivalence of a quadratic and a linear model defined by \eqref{ex4}. The threshold in \eqref{h0inf} is chosen as $\epsilon_\infty=1$. The numbers in brackets show the simulated power of the test proposed in \cite{gsteiger2011}.}
}
\end{table}

\begin{example} {\bf  {($\mathcal{E} = [0,4]  $)}} \label{exam4} {\rm
We conclude this section with an investigation of the models in \eqref{ex2} which represents somehow the extreme case, as the set of extremal points
of the true absolute difference is given by $\mathcal{E}= [0,4]$, which is the entire dose range. In Table
\ref{tab9}  we display the rejection probabilities of the bootstrap  test \eqref{testinf}
under the null hypothesis.
 Corresponding results under the alternative are shown in Table \ref{tab10}, where it is demonstrated that  the bootstrap test
  \eqref{testinf} yields again a substantial improvement in power compared to the test of \cite{gsteiger2011}. While this test has practically no power, the new bootstrap test proposed in this paper is able to establish equivalence between the curves with reasonable Type II error rates, if the total sample size is larger than $50$.
  }
\end{example}

%\FloatBarrier
\begin{table}[!h]
  {\tiny
\centering
\begin{tabular}{||c|c|ccc||ccc||}
\hline
\multicolumn{1}{||c|}{} &   \multicolumn{1}{|c|}{} & \multicolumn{3}{|c||}{$\alpha=0.05$} & \multicolumn{3}{|c||}{$\alpha=0.1$} \\\hline
\multicolumn{1}{||c|}{}   & \multicolumn{1}{|c|}{} &
\multicolumn{3}{|c||}{$(\sigma_1^2,\sigma_2^2)$} & \multicolumn{3}{|c||}{$(\sigma_1^2,\sigma_2^2)$} \\\hline
$(n_1,n_2)$   & $d= d_\infty$ & $(0.25,0.25)$ & $(0.5,0.5)$ & $(0.25,0.5)$ &
$(0.25,0.25)$ & $(0.5,0.5)$ & $(0.25,0.5)$
  \\ \hline
$(10,10)$ & 1    & 0.000 (0.000) & 0.004 (0.000) & 0.001 (0.000)  & 0.007 (0.000) & 0.019 (0.000) &0.010 (0.000) \\
$(10,10)$ & 0.75 & 0.000 (0.000) & 0.008 (0.000) & 0.006 (0.000) & 0.013 (0.002) & 0.041 (0.000) & 0.020 (0.000)  \\
$(10,10)$ & 0.5  & 0.015 (0.001) & 0.040 (0.000) & 0.016 (0.000)  & 0.050 (0.005) & 0.104 (0.000) & 0.054 (0.002) \\
\hline
$ (10,20)$ & 1   & 0. 000 (0.000)& 0.000 (0.000) & 0.000 (0.000)  & 0.005 (0.000) & 0.002 (0.000) &0.003 (0.000) \\
$ (10,20)$ & 0.75&  0.001 (0.000)& 0.004 (0.000) & 0.000 (0.000)  & 0.005 (0.000) & 0.023 (0.000) &0.006 (0.000)  \\
$ (10,20)$ & 0.5 &  0.018 (0.000)& 0.016 (0.000) &  0.012 (0.000) & 0.045 (0.000) & 0.051 (0.000) &0.037 (0.000) \\
\hline
$(20,20)$ & 1    & 0.000 (0.000) & 0.000 (0.000) & 0.000 (0.000) & 0.000 (0.000) & 0.004 (0.000) & 0.006 (0.000)  \\
$(20,20)$ & 0.75 & 0.000 (0.000) & 0.002 (0.000) & 0.000 (0.000)  & 0.003 (0.000) & 0.010 (0.002) & 0.002 (0.000) \\
$(20,20)$ & 0.5  & 0.006 (0.001) & 0.019 (0.000) & 0.016 (0.000) & 0.027 (0.001) & 0.051 (0.000) &0.046 (0.000)  \\
\hline
$(50,50)$ & 1    & 0.000 (0.000) & 0.000 (0.000) & 0.000 (0.000) & 0.000 (0.000) & 0.000 (0.000) & 0.001 (0.000)   \\
$(50,50)$ & 0.75 & 0.006 (0.000) & 0.000 (0.000) & 0.000 (0.000) &0.004 (0.000)  & 0.007 (0.000) &0.002 (0.000)  \\
$(50,50)$ & 0.5  & 0.003 (0.000) & 0.005  (0.000)& 0.004  (0.000)& 0.018 (0.000) & 0.027 (0.000) & 0.034 (0.000)  \\
\hline
\hline
\end{tabular}
\caption{\label{tab9} \it Simulated Type I error rates  of the bootstrap $d_\infty$-test
 \eqref{testinf}  for the equivalence of two shifted EMAX-models defined by \eqref{ex2}.  The threshold in \eqref{h0inf} is chosen as $\epsilon_\infty=0.5$. The numbers in brackets show the simulated Type I error  rates of the test proposed by \cite{gsteiger2011}.}
}
\end{table}

 \begin{table}[!h]
  {\tiny
\centering
\begin{tabular}{||c|c|ccc||ccc||}
\hline
\multicolumn{1}{||c|}{} &   \multicolumn{1}{|c|}{} & \multicolumn{3}{|c||}{$\alpha=0.05$} & \multicolumn{3}{|c||}{$\alpha=0.1$} \\\hline
\multicolumn{1}{||c|}{}   & \multicolumn{1}{|c|}{} &
\multicolumn{3}{|c||}{$(\sigma_1^2,\sigma_2^2)$} & \multicolumn{3}{|c||}{$(\sigma_1^2,\sigma_2^2)$} \\\hline
$(n_1,n_2)$   & $d= d_\infty$ & $(0.25,0.25)$ & $(0.5,0.5)$ & $(0.25,0.5)$ &
$(0.25,0.25)$ & $(0.5,0.5)$ & $(0.25,0.5)$
  \\ \hline
$(10,10)$ & 0.25  & 0.062 (0.000)& 0.050 (0.000)& 0.053 (0.000) & 0.147 (0.000)& 0.118 (0.000)& 0.118 (0.000) \\
$(10,10)$ & 0.1   & 0.100 (0.000)& 0.070 (0.000)& 0.099 (0.000) & 0.195 (0.000)& 0.137 (0.000)& 0.190 (0.000) \\
$(10,10)$ & 0     & 0.109 (0.000)& 0.090 (0.000)& 0.092 (0.000)& 0.216 (0.000)& 0.143 (0.000)& 0.176 (0.000) \\
\hline
$ (10,20)$ & 0.25 & 0.077 (0.000)& 0.077 (0.000)& 0.074 (0.000) & 0.157 (0.000)& 0.142 (0.000)& 0.141 (0.000) \\
$ (10,20)$ & 0.1  & 0.118 (0.001)& 0.077 (0.001)&  0.100 (0.000) & 0.227 (0.002)& 0.163 (0.002)& 0.176 (0.000) \\
$ (10,20)$ & 0    & 0.151 (0.001)& 0.078 (0.001)& 0.118 (0.000)&0.275  (0.004)&0.165 (0.003) & 0.213 (0.000)  \\
\hline
$(20,20)$ & 0.25  & 0.085 (0.000)& 0.060 (0.000)& 0.076  (0.000) & 0.171 (0.005)& 0.134 (0.001)& 0.162 (0.000) \\
$(20,20)$ & 0.1   & 0.158 (0.000)& 0.090 (0.000)& 0.112 (0.000) & 0.309 (0.007)& 0.184 (0.002)& 0.220 (0.001)  \\
$(20,20)$ & 0     & 0.178 (0.003)& 0.108 (0.001)& 0.120 (0.003) & 0.324 (0.013)& 0.209 (0.001)& 0.219 (0.008) \\
\hline
$(50,50)$ & 0.25  & 0.162 (0.023)& 0.086 (0.000)& 0.098 (0.006) & 0.283 (0.084)& 0.178 (0.007)& 0.218 (0.034) \\
$(50,50)$ & 0.1   & 0.390 (0.117)& 0.212 (0.002)& 0.232 (0.017)& 0.568 (0.325)& 0.349 (0.018)& 0.398 (0.101) \\
$(50,50)$ & 0     & 0.457 (0.157)& 0.211 (0.012)& 0.266 (0.032) & 0.630 (0.364)&0.363 (0.033) & 0.438 (0.172) \\
\hline\hline
\end{tabular}
\caption{\label{tab10} \it Simulated power of the bootstrap $d_\infty$-test
 \eqref{testinf}  for the equivalence of two shifted EMAX-models defined by \eqref{ex2}. The threshold in \eqref{h0inf} is chosen as $\epsilon_\infty=0.5$. The numbers in brackets show the simulated power of the test proposed in \cite{gsteiger2011}.}
}
\end{table}
%\FloatBarrier

\FloatBarrier

\section{Case study} \label{sec6}
\def\theequation{6.\arabic{equation}}
\setcounter{equation}{0}

\begin{figure}[h]
	\centering
 		\includegraphics[width=0.4\textwidth]{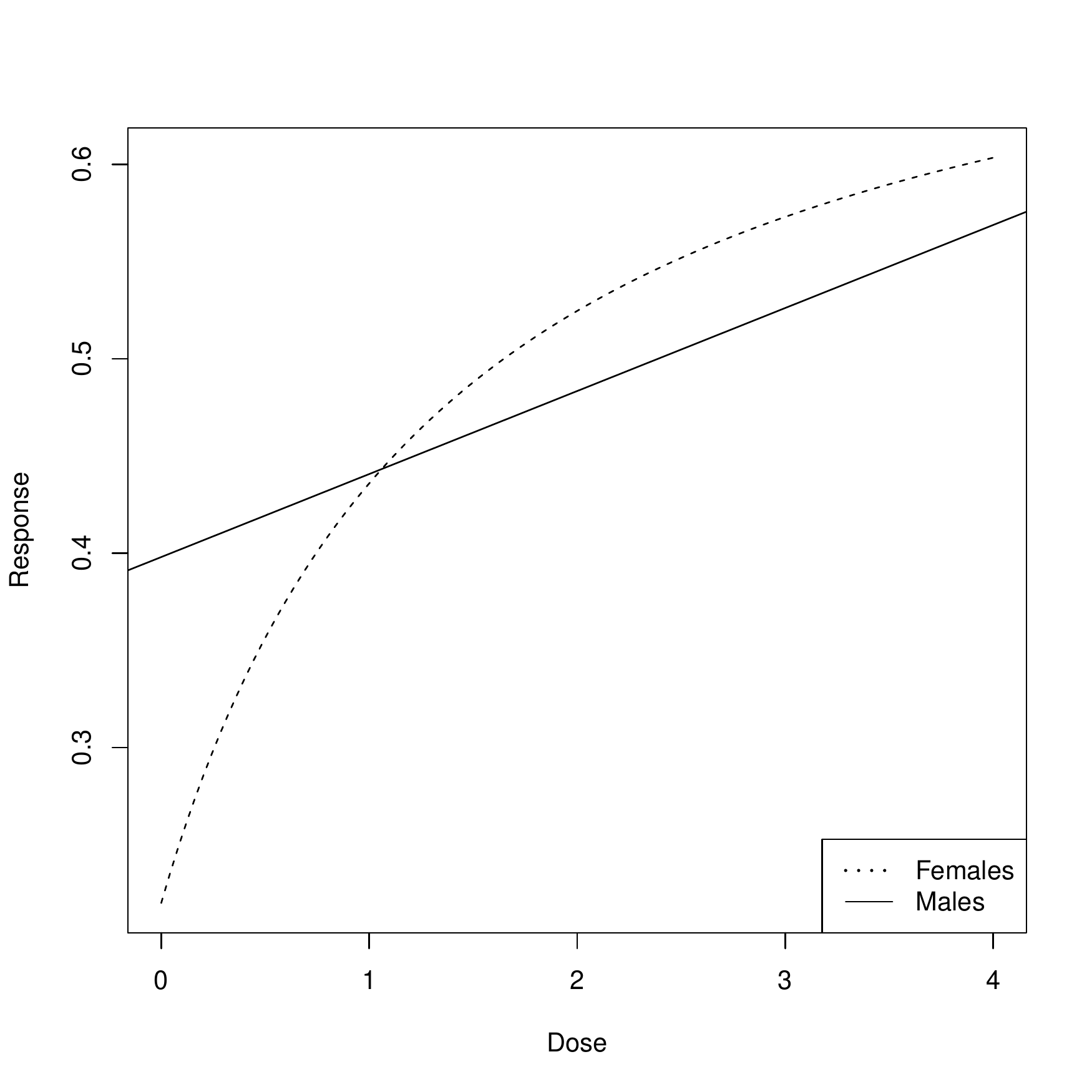}  ~~~
		\includegraphics[width=0.4\textwidth]{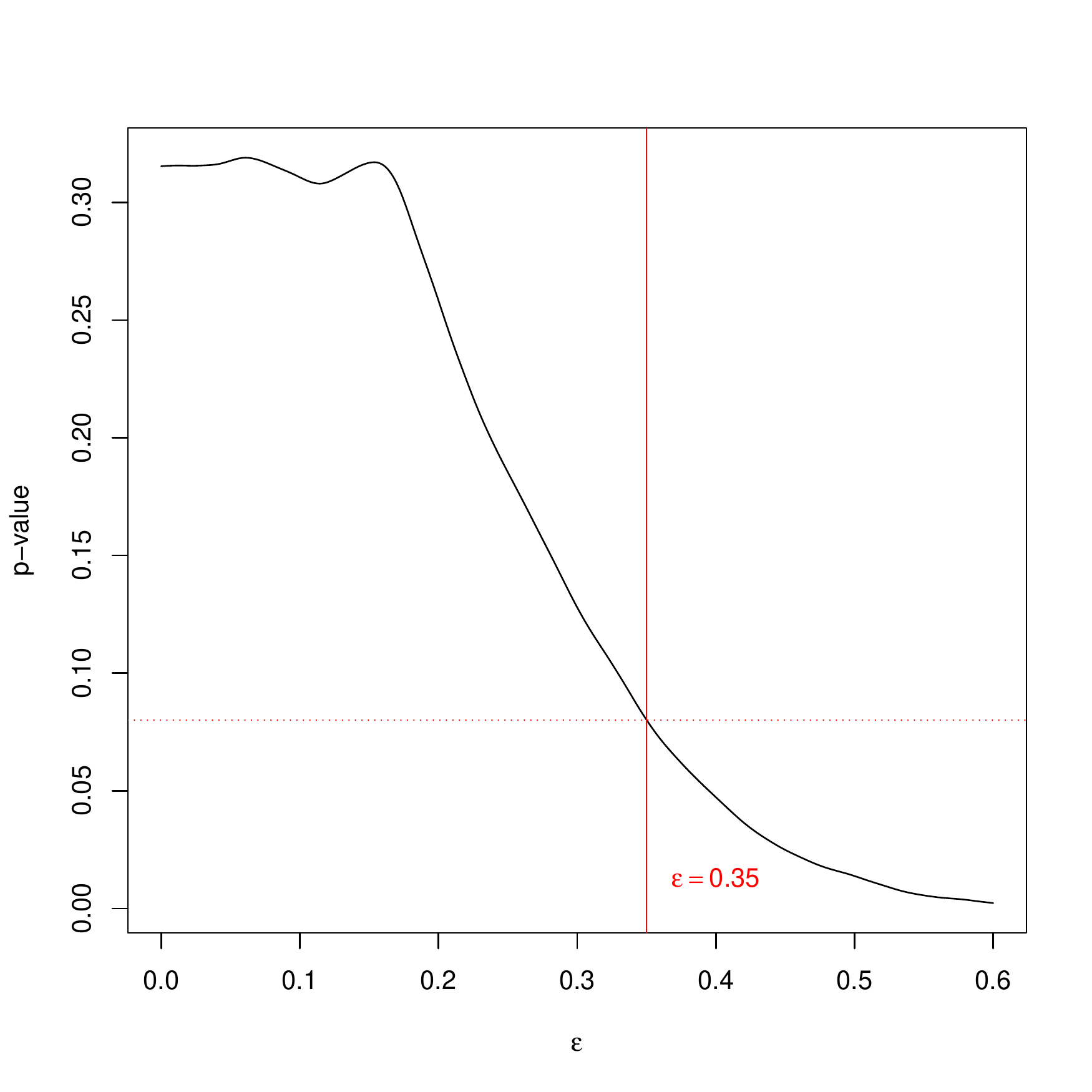}
	\caption{\it Left panel: Fitted dose response curves for male (linear model) and female (Emax model) patients
	from  the IBS case study. Right panel:  $p$-values of the test \eqref{testinf} for different values of the threshold
	$\varepsilon_\infty$ in the hypothesis \eqref{h0inf}.}	
	\label{fig1}
\end{figure}

In this section we illustrate the new methodology with the dose finding study described in \cite{biesheuvel2002many}.
Female and male patients with Irritable Bowel Syndrome (IBS)
were randomized to one of the five doses 0 (placebo), 1, 2, 3, and 4.
We use the blinded dose levels for confidentiality.
The primary endpoint was a baseline adjusted abdominal pain score
with larger values corresponding to a better treatment effect.
In total, $369$ patients completed the study,
with nearly balanced allocation across the five doses.
The data is available in the \texttt{R} package
\texttt{DoseFinding} from \cite{borpinbre2010}. \\
For this example, we used the linear model
$m_1(x,\beta_1) = \beta_{1,1} + \beta_{1,2}x$ for the males
and the Emax model
$m_2(x,\beta_2) = \beta_{2,1} + \beta_{2,2}\frac{x}{\beta_{2,3} + x}$ for the females.
Note that males and females were assigned to the same set of doses.
The estimators in the linear and in the Emax model are given by $\hat{\beta}_{1} = (0.398, 0.043)$
and $\hat{\beta}_{2} = (0.220, 0.517, 1.396)$, respectively.
The left part of Figure \ref{fig1} displays the fitted dose response
models for both groups in the interval $ [0, 4]$.  As it can also be observed from Figure~\ref{fig1}, the maximum
distance between the two curves is $\hat{\beta}_{1,1} - \hat{\beta}_{2,1} = 0.1784$, attained at $x=0$.

\begin{table}[h]
  {\tiny
\center
		\begin{tabular}{|c|c|c|}
		\hline
		$\epsilon_\infty$ & $\alpha=0.05$ & $\alpha=0.1$\\
		\hline
		0.3 & 0.1293 & 0.1628\\
		0.35 & 0.1578 & 0.1972\\
		0.4 & 0.1867 & 0.2322\\
		\hline
		\end{tabular}
		\caption{\it
		\label{tab_quant}
		Quantiles of  the bootstrap test \eqref{testinf}  (from $5000$  replications) in the IBS case study  for different values of the threshold
	$\varepsilon_\infty$ in the hypothesis \eqref{h0inf}.}
	}
\end{table}
\noindent
We first compare males and females with respect to the maximal deviation distance  $d_\infty$ defined in \eqref{sup}.
For this purpose we apply the bootstrap test \eqref{testinf} proposed in Section~\ref{sec4}, which is implemented
with the \texttt{R} package \texttt{TestingSimilarity} from \cite{moellenhoff2015}. In Table \ref{tab_quant}
we display the quantiles  of  the bootstrap test \eqref{testinf}    for different values of the threshold $\varepsilon_\infty$
in the hypothesis \eqref{h0inf}. These values  are calculated  by  $5000$ bootstrap replications. For example, if  $\epsilon_\infty=0.35$ we
obtain the quantiles $q_{0.1,\infty } = 0.1972$ for $\alpha=0.1$ and
 $q_{0.05,\infty } = 0.1578$ for $\alpha=0.05$, while the value of the test statistic is given by $d_\infty(\hat \beta_1, \hat \beta_2) = 0.1784$. Thus, we can reject the null hypothesis \eqref{h0inf} at the significance level
 $\alpha=0.1$, but not at $\alpha=0.05$. In the right part of  Figure \ref{fig1} we display  the $p$-value
 $$\frac{1}{B} \sum_{i=1}^B I (d_\infty^{*(i)}\leq \hat d_\infty)
 $$
  of the bootstrap test \eqref{testinf} as a function of  the threshold $\epsilon_\infty$. We observe that the $p$-value corresponding to the
   choice  $\epsilon_\infty=0.35$ is given by $0.078$.
 The plateau of the curve is a consequence of the special construction of the test. If $\epsilon_\infty\leq d_\infty=0.1784$,
  the constrained estimators in \eqref{MLcons} coincide with the unconstrained  estimators
  and consequently the bootstrap data is generated with the same parameters for each of these different values of
  $\epsilon_\infty$.

In comparison, using the test of \cite{gsteiger2011} discussed in Section \ref{sec5}, the
maximum (minimum) of the upper (lower) confidence band is given by $0.282$ ($-0.450$) and $0.227$ $(-0.390)$ for $\alpha=0.05$ and $\alpha=0.1$ respectively. Thus, the null hypothesis cannot be rejected at either of these two significance levels as the bands are not completely contained in the rectangle $\mathcal{X} \times [-\epsilon_\infty,\epsilon_\infty]$. Therefore, similarity of the curves at a significance level of $\alpha=0.1$ can only be claimed if $\epsilon_\infty$ is larger than $0.390$. This illustrates again that the test proposed in \cite{gsteiger2011} is conservative.\\
It might be also of interest to compare both curves with respect to the squared  $L^2-$distance \eqref{ell2}, which is
given by  $d_2(\hat \beta_1,\hat \beta_2) =0.0126$ for the IBS data. The $0.05$ and $0.1$ quantile of the bootstrap
distribution are given by  $q_{0.05,2} = 0.0108$ and $q_{0.1,2} = 0.0169$, respectively, for a choice of $\epsilon_2=0.05$. Thus, we can reject the null hypothesis
\eqref{hypothesesL2} at  significance level  $\alpha=0.1$  but not at $\alpha=0.05$.
In this case  the $p$-value  is given by $0.059$.
Finally, we illustrate the application of the asymptotic test \eqref{testl2asy}. The variance estimator defined
 in \eqref{sigmat} is  given by $\hat \sigma_{d_2}^2=0.0010$ and for  $\epsilon_2=0.05$ we obtain the critical values
 $-0.0022$ ($\alpha=0.05$) and $ 0.0093$  ($\alpha=0.1$) for the test \eqref{testl2asy}. Therefore the null hypothesis cannot
 be rejected at either of these two significance levels and the corresponding $p$-value is given by $0.1193$.
 These findings coincide with the results  of the simulation study, which shows  that the
  bootstrap test \eqref{test} is more accurate.
\medskip
\medskip

{\bf Acknowledgements}
This work has been supported in part by the
Collaborative Research Center "Statistical modeling of nonlinear
dynamic processes" (SFB 823, Project C1) of the German Research Foundation (DFG).
Kathrin M÷llenhoff's research has received funding from the European Union Seventh Framework Programme [FP7 2007û2013] under grant agreement no 602552 (IDEAL - Integrated Design and Analysis
of small population group trials).
The authors would like to thank Martina
Stein, who typed parts of this manuscript with considerable
technical expertise.
 We are very grateful to two  referees, the associate editor and the editor for their
 constructive comments, which led to substantial improvement of  an earlier version of
 this manuscript.

%\end{appendix}
 \bibliographystyle{apalike}
\setlength{\bibsep}{1pt}
\begin{small}
\bibliography{curves}
 \end{small}

\section{Appendix: Technical details}  \label{sec7}
\def\theequation{7.\arabic{equation}}
\setcounter{equation}{0}

The theoretical results of this paper are proved under the following assumptions.
\begin{assumption} \label{2.0}
The errors $\eta_{\ell,i,j}$ are independent, have finite variance $\sigma_\ell^2$ and expectation zero.
\end{assumption}
\begin{assumption} \label{2.1}
The covariate region $\mathcal{X} \subset \mathbb{R}^d$ is compact and the number and location of dose levels $k_\ell$ does not depend on $n_\ell,\ \ell=1,2$.
\end{assumption}
\begin{assumption}\label{2.2}
All estimators of the parameters $\beta_1, \beta_2$ are computed over compact sets $B_1 \subset \mathbb{R}^{p_1}$ and $B_2 \subset \mathbb{R}^{p_2}$.
\end{assumption}
\begin{assumption}\label{2.3}
The regression functions $m_1$ and $m_2$ are twice continuously differentiable with respect to the parameters for all $b_1,b_2$ in neighbourhoods of the true parameters $\beta_1,\beta_2$ and all $x \in \mathcal{X}$. The functions $(x,b_\ell) \mapsto m_\ell(x,b_\ell)$ and their first two derivatives are continuous on $\mathcal{X} \times B_\ell$.
\end{assumption}
%\begin{assumption} \label{2.4}
%The gradients with respect to the parameters are uniformly bounded, that is
%$$\sup_{x\in \mathcal{X}}\left\|\tfrac{\partial }{\partial\beta_\ell}  m_\ell(x,\beta_\ell)\right\|<\infty,\ \ell=1,2.$$
%\end{assumption}
\begin{assumption} \label{2.5}
Defining
$\psi_{a,\ell}^{(n)}(b) : = \sum_{i=1}^{k_\ell} \frac{n_{\ell,i}}{n_\ell} (m_\ell(x_{\ell,i},a) - m_\ell(x_{\ell,i},b))^2,
$
we assume that for any $u>0$ there exists a constant $v_{u,\ell}>0$ such that
$$
\liminf_{n_1,n_2 \to \infty} \inf_{a \in B_\ell}\inf_{|b-a| \geq u} \psi_{a,\ell}^{(n)}(b) \geq v_{u,\ell} \qquad \ell=1,2.
$$
\end{assumption}
In particular, under Assumptions \ref{2.0} - \ref{2.5} the least squares estimator can be linearized.
To be precise, consider arbitrary sequences $(\beta_{\ell,n})_{n\in\mathbb{N}}$ and $(\sigma_{\ell,n})_{n\in\mathbb{N}}$ in $B_\ell$ and $\mathbb{R}^+$ such that $\beta_{\ell,n} \to \beta_\ell$ and $\sigma_{\ell,n} \to \sigma_\ell>0$
as $n_1,n_2 \to \infty $  ($\ell = 1,2$) and denote by $Y_{\ell,i,j}^{(n)}$ data of the form given in \eqref{mod1a}
%, \eqref{mod1b}
with $\beta_\ell$ replaced by $\beta_{\ell,n}$ and $\eta_{\ell,i,j}$ independent and identically distributed (for each fixed $n$) with mean $0$ and finite variances $\sigma_{\ell,n}^2$. Then the least squares estimators $\hat \beta_\ell^{(n)}$ computed from $Y_{\ell,i,j}^{(n)}$ satisfy
\begin{equation} \label{exp_tri}
\sqrt{n_\ell} \ (\hat \beta_\ell^{(n)} - \beta_{\ell,n}) = \frac {1}{\sqrt{n_\ell}} \sum^{k_\ell}_{i=1} \sum^{n_{\ell,i}}_{j=1} \phi_{\ell,i,j}+ o_{\mathbb{P}}(1), \qquad \ell=1,2,
\end{equation}
where the functions $\phi_{\ell,i,j}^{(n)}$ are given by
\begin{equation} \label{phij_tri}
\phi_{\ell,i,j} = \tfrac {\eta_{\ell,i,j}}{\sigma^2_{\ell,n}}\Sigma^{-1}_{\ell,n}  \tfrac{\partial }{\partial b_l}  m_\ell(x_{\ell,i},b_\ell)\big|_{b_l = \beta_{\ell,n}}, \qquad \ell=1,2,
\end{equation}
and $\Sigma_{\ell,n}$ takes the form
\begin{equation}\label{sigmal_tri}
\Sigma_{\ell,n} ={1 \over \sigma_{\ell,n}^2} \sum_{i=1}^{k_\ell}\zeta_{\ell,i}\tfrac{\partial }{\partial b_l}  m_\ell(x_{\ell,i},b_\ell)\big|_{b_l = \beta_{\ell,n}}
 \big(\tfrac{\partial }{\partial b_l}  m_\ell(x_{\ell,i},b_\ell)\big|_{b_l = \beta_{\ell,n}} \big)^T~,~~\ell=1,2.
\end{equation}

\paragraph{Proof of Theorem \ref{thm1}:} \label{proofthm1}
Let $\ell_\infty (\mathcal{X})$ denote the space of all bounded real valued functions of the form $g: \mathcal{X} \to  \er$. The mapping $\Phi:  \mathbb{R}^{p_1+p_2}\rightarrow\ell_\infty (\mathcal{X}) $
defined by
\begin{equation} \label{map}
 (\theta_1,\theta_2 )\mapsto \Phi(\theta_1,\theta_2):
 \left\{
 \begin{array}{l}
 \mathcal{X} \mapsto  \er \\
 x \mapsto  \big(\tfrac{\partial }{\partial b_1}  m_1(x,b_1)\big|_{b_1 = \beta_1} \big)^T\theta_1 -\big(\tfrac{\partial }{\partial b_2}  m_2(x,b_2)\big|_{b_2 = \beta_2} \big)^T \theta_2
\end{array}
\right.
\end{equation}
 {is continuous due to Assumptions \ref{2.1}-\ref{2.3}}, where we use the Euclidean and the supremum norm on
 $\mathbb{R}^{p_1+p_2}$ and $\ell_\infty (\mathcal{X})$, respectively.
Consequently, the continuous mapping theorem  [see \cite{vaart1998}]  and \eqref{MLasy} yield that the process
\begin{eqnarray*}
\{ \sqrt{n} G_n(x) \}_{x\in \mathcal{X}} ~:=
\big\{\sqrt{n}\big((\tfrac{\partial }{\partial b_1}  m_1(x,b_1)\big|_{b_1 = \beta_1})^T(\hat{\beta_1}-\beta_1)-(\tfrac{\partial }{\partial b_2}  m_2(x,b_2)\big|_{b_2 = \beta_2})^T
(\hat{\beta_2}-\beta_2)\big)
\big\}_{x\in \mathcal{X}}
\end{eqnarray*}
converges weakly to a centered  Gaussian process $\{ G(x) \}_{x\in \mathcal{X}} $ in  $\ell_\infty (\mathcal{X})$, which is defined
by \begin{equation}
G(x)=\big(\tfrac{\partial }{\partial b_1}  m_1(x,b_1)\big|_{b_1 = \beta_1}\big)^T \sqrt{\lambda}\Sigma_1^{-1/2}
Z_1-\big(\tfrac{\partial }{\partial b_2}  m_2(x,b_2)\big|_{b_2 = \beta_2}\big)^T\sqrt{\tfrac{\lambda}{\lambda-1}}
\Sigma_2^{-1/2} Z_2,\label{G}\end{equation}
where $Z_1$ and $Z_2$ are independent $p_1$- and $p_2$-dimensional  standard normal distributed random variables, respectively,
i.e. $Z_\ell \sim {\cal N} (0, I_{p_\ell})$, $\ell=1,2$.  A straightforward calculation shows that the  covariance kernel
of the process $\{ G(x) \}_{x\in \mathcal{X}} $ is given by \eqref{kernel}.  Now a Taylor expansion gives
\begin{eqnarray} \label{delta}
 p_n (x) := \big(m_1(x,\hat{\beta}_1 )-m_1 (x,\beta_1 )\big) -\big(m_2(x,\hat{\beta}_2 )-m_2 (x,\beta_2 )\big)
% \sum_{\ell=1}^2 \big(m_\ell(x,\hat{\beta}_\ell )-m_\ell (x,\beta_\ell )\big)
= G_n(x) +o_{\mathbb{P}}\Big(\sqrt{\frac{1}{n}}\Big),
\end{eqnarray}
uniformly with respect to $x \in \mathcal{X}$, and it therefore follows that
\begin{equation} \label{weakd}
 \{ \sqrt{n} p_n(x) \}_{x\in \mathcal{X}} \stackrel{\mathcal{D}}{\longrightarrow}\left\{G(x)\right\}_{x\in \mathcal{X}}.
 \end{equation}
Recalling the definition of $\Delta(x,\beta_1,\beta_2)$ in \eqref{truediff},
observing the representation
\begin{eqnarray*}
\sqrt{n}(\hat{d}_2-d_2)&=& \sqrt{n} \Big(  \int_\mathcal{X} \Delta^2(x,\hat \beta_1, \hat \beta_2)    dx  -\int_\mathcal{X} \Delta^2(x,  \beta_1,   \beta_2) dx \Big) \\
%&=&
%\sqrt{n}\int_\mathcal{X} \big (p_n(x)+2 \Delta(x, \beta_1,  \beta_2) \big) p_n(x)dx\\
&=&\int_\mathcal{X} \sqrt{n} p_n^2(x) dx + 2\sqrt{n}\int_\mathcal{X}\Delta(x, \beta_1,  \beta_2) p_n(x)dx,
\end{eqnarray*}
and from the continuous mapping theorem we    obtain
$\sqrt{n}(\hat{d}_2-d_2)\stackrel{\mathcal{D}}{\rightarrow}2\int_\mathcal{X}\Delta(x,  \beta_1,  \beta_2) G(x)dx$,
where $G$ denotes the Gaussian process defined in \eqref{G}.
Now it is easy to see that the distribution on the right-hand side is a centered normal distribution with
variance $\sigma^2_{d_2}$ defined in \eqref{sigmat}. This completes the proof of Theorem \ref{thm1}. \hfill $\Box$

\paragraph{Proof of Theorem \ref{thm2}:}  \label{A.1A}~~\\
\textit{Proof of (1).}
First we will determine the asymptotic distribution of the bootstrap estimators $\hat \beta^*_1$ and $\hat \beta^*_2$. Then we
use similar arguments as given in the proof of Theorem \ref{thm1} to derive the asymptotic distribution  of the statistic $\hat d_2^*$
(appropriately standardized). Finally, in a third step, we establish the statement  $\eqref{level}$.\\
 Recall  the definition of the estimators in \eqref{MLcons} and note
 that it follows from Assumptions \ref{2.0}-\ref{2.5} that $ d_2 \geq \varepsilon_2$ under  the null hypothesis . %the consistency of the unconstrained estimates that $\hat d_2 \stackrel{\mathbb{P}}{\longrightarrow} d_2 \geq \varepsilon$ implies
We distinguish two cases. If $d_2 > \varepsilon_2$, consistency of $\hat \beta_\ell$ implies that $\hat d_2 > \varepsilon_2$ with probability tending to one, and thus $\hat{\hat{\beta_\ell}} = \hat \beta_\ell$ with probability tending to one. Next consider the case $d_2 = \varepsilon_2$. Let $M := \{(b_1,b_2) \in B_1\times B_2: d_2(b_1,b_2) = \varepsilon_2 \}$ and note that $(\beta_1,\beta_2) \in M$. Define
\begin{eqnarray*}
\psi_{\ell}^{(n)}(b) &:=& \sum_{i=1}^{k_\ell} \frac{n_{\ell,i}}{n_\ell} (m_\ell(x_{\ell,i},\beta_\ell) - m_\ell(x_{\ell,i},b))^2~, ~
\hat \gamma_\ell := \frac{1}{n_\ell}\sum_{i=1}^{k_\ell}\sum_{j=1}^{n_{\ell,i}} (\eta_{\ell,i,j})^2
\\
\hat \psi_{\ell}(b) &:=& \frac{1}{n_\ell}\sum_{i=1}^{k_\ell}\sum_{j=1}^{n_{\ell,i}} (m_\ell(x_{\ell,i},\beta_\ell) + \eta_{\ell,i,j} - m_\ell(x_{\ell,i},b))^2.
\end{eqnarray*}
By definition (see  \eqref{MLcons} and \eqref{constr}), we have
$
(\tilde \beta_1,\tilde\beta_2) = \arg \min_{(b_1,b_2) \in M} \sum_{\ell=1}^2 \hat \psi_{\ell}(b_\ell).
$
Moreover,
\[
\arg\min_{(b_1,b_2) \in M } \sum_{\ell=1}^2 \hat \psi_{\ell}(b_\ell) = \arg\min_{(b_1,b_2) \in M } \sum_{\ell=1}^2 (\hat \psi_{\ell}(b_\ell) - \hat\gamma_\ell),
\]
and
\[
\hat \psi_{\ell}(b_\ell) - \hat\gamma_\ell = \psi_{\ell}^{(n)}(b) + 2\sum_{i=1}^{k_\ell}(m_\ell(x_{\ell,i},\beta_\ell)- m_\ell(x_{\ell,i},b)) \frac{1}{n_{\ell}}\sum_{j=1}^{n_{\ell,i}}  \eta_{\ell,i,j}.
\]
{Observing that the terms $|m_\ell(x_{\ell,i},\beta_\ell)- m_\ell(x_{\ell,i},b)| $} are uniformly bounded (with respect to $b \in B_\ell$ and $ x_{\ell,i} \in {\cal X}$) it follows that
\[
R_n := \sum_{\ell=1}^2\sup_{b \in B_\ell} \Big| 2\sum_{i=1}^{k_\ell}(m_\ell(x_{\ell,i},\beta_\ell)- m_\ell(x_{\ell,i},b)) \frac{1}{n_{\ell}}\sum_{j=1}^{n_{1,\ell}}  \eta_{\ell,i,j} \Big| = o_\mathbb{P}(1)
\]
since $\max_{i=1,...,k_1}|\frac{1}{n_{\ell}}\sum_{j=1}^{n_{\ell,i}} \eta_{\ell,i,j}| = o_\mathbb{P}(1)$. By similar arguments as given after \eqref{eq:helpkons}, we obtain $|\tilde\beta_\ell - \beta_{\ell}| = o_\mathbb{P}(1)$, $\ell =1,2$ (recall that $(\beta_1,\beta_2) \in M$). Since for $\ell = 1,2$ we have $|\hat{\hat{\beta_\ell}} - \beta_{\ell}| \leq |\hat\beta_\ell - \beta_{\ell}| + |\tilde\beta_\ell - \beta_{\ell}|$ and   it follows from consistency of $\hat\beta_\ell$ that

\begin{equation}\label{consconstr}
{\hat{\hat{\beta}}_{\ell}} \stackrel{\mathbb{P}}{\longrightarrow} \beta_\ell \quad \ell=1,2, \quad \mbox{whenever} \quad d_2 \geq \varepsilon_2.
\end{equation}

For $\ell=1,2$ let ${\cal Y} _\ell =\sigma (  Y_{\ell,i,j} | i=1,\ldots ,k_\ell, j=1,\ldots , n_{\ell ,i} )$ denote the $\sigma$-field generated by the random variables $ \{Y_{\ell, i,j}| i=1,\ldots ,k_\ell, j=1,\ldots , n_{\ell ,i}\} $ and ${\cal Y} := \sigma({\cal Y} _1,{\cal Y} _2)$ (note that we do not display the dependence of these quantities on the sample size). Given \eqref{consconstr} and the consistency of $\hat \sigma_\ell$, the discussion after Assumption \ref{2.5} yields
\[
\sqrt{n_\ell}(\hat \beta_\ell^*-\hat{\hat{\beta}}_\ell)
=  \sum_{i=1}^{k_\ell}  \frac {1}{  \hat \sigma_\ell}\hat{\hat{\Sigma}}^ {-1}_\ell  \tfrac {\partial}{\partial \beta_\ell}  m_\ell (x_{\ell,i,}, \beta_\ell )\big|_{\beta_\ell=\hat{\hat{\beta}}_\ell} \frac{1}{\sqrt{n_\ell}} \sum_{j=1}^{n_{\ell,i}}\tfrac {\eta^*_{\ell,i,j}}{  \hat \sigma_\ell}  ~+ ~o_{\mathbb{P}}(1)~,~~\ell=1,2.
\]
where  the $p_1 \times p_1$ and $p_2\times p_2$ dimensional  matrices $\hat{\hat{\Sigma}}^ {-1}_1$ and $\hat{\hat{\Sigma}}^ {-1}_2$  are defined by
  $$
  \hat{\hat{\Sigma}}_\ell = \frac{1}{  \hat \sigma^2_\ell}   \sum^{k_\ell}_{i=1}
  \zeta_{\ell,i}
  \big ( \tfrac {\partial}{\partial \beta_\ell}  m_\ell (x_{\ell,i,}, \beta_\ell )\big|_{\beta_\ell=\hat{\hat{\beta}}_\ell} \big) \big( \tfrac {\partial}{\partial \beta_\ell}  m_\ell (x_{\ell,i,}, \beta_\ell )\big|_{\beta_\ell=\hat{\hat{\beta}}_\ell} \big)^T.
  $$
Since by construction the $\frac {\eta^*_{\ell,i,j}}{  \hat \sigma_\ell}$ are i.i.d. with unit variance and independent of ${\cal Y} $, the classical central limit theorem implies that, conditionally on ${\cal Y} $ in probability
\begin{equation} \label{eq:norm}
\sqrt{n_\ell}\cdot \Sigma_\ell ^{\frac{1}{2}}(\hat \beta^*_\ell-\hat{\hat{\beta_\ell }})\overset{D}\longrightarrow \mathcal{N}(0,I_{p_\ell})~, \qquad  \ell=1,2,
\end{equation}
where the matrix $\Sigma_\ell$ is defined in \eqref{sigmal}.
Observing the definition  of the    statistic
$$\hat d^*_2 = d_2 (\hat \beta_1^*, \hat \beta_2^*)=\int_\mathcal{X}(m_1(x,\hat \beta_1^*)-m_2(x,\hat\beta^*_2))^2 dx,$$
it now follows by
the same arguments as given in the proof of Theorem \ref{thm1}  that\begin{equation}
\frac{\sqrt{n}}{\sigma_{d_2}} (\hat d_2^*-\hat{\hat{d_2}})
\stackrel{\mathcal{D}}{\rightarrow}\mathcal{N}(0,1)\label{bootstrapkonv}
\end{equation}
conditionally on ${\cal Y} $ in probability. Now recall that  $\hat{q}_{\alpha,2}$ is the $\alpha$-quantile of the bootstrap statistic  $\hat d^*_2$ conditionally on ${\cal Y} $ and note that, almost surely,
\be \label{bootlevel}
 \alpha =
\mathbb{P} (\hat d^*_2<\hat{q}_{\alpha,2} |~{\cal Y}  )=\mathbb{P}\Big(\frac{\sqrt{n}(\hat d^*_2-\hat{\hat{d}}_2)}
{\sigma_{d_2}}<\frac{\sqrt{n}(\hat{q}_{\alpha,2}-\hat{\hat{d}}_2)}{\sigma_{d_2}}\Big| ~{\cal Y} \Big).
\ee
Letting  $ \hat{p}_{\alpha}:= {\sqrt{n}(\hat{q}_{\alpha,2}-\hat{\hat{d}}_2)}/{\sigma_{d_2}} $
%denotes  the $\alpha$-quantile of the distribution of the random variable $\frac{\sqrt{n}(d^*_2-\hat{\hat{d}}_2)}{\hat \sigma_{d_2}}$ conditionally on ${\cal Y} $, then
it follows  from $\eqref{bootstrapkonv}$, \eqref{bootlevel} and {Lemma 21.2} in  \cite{vaart1998}   that
\begin{equation}\hat{p}_{\alpha}\stackrel{\mathbb{P}}\longrightarrow u_{\alpha},\label{quantile2}\end{equation}
where $u_{\alpha}$ denotes the $\alpha$-quantile of the standard normal distribution.
This relation implies for  any $\alpha<0.5$  that
\begin{equation}\lim_{n\rightarrow \infty}\mathbb{P} (\hat{q}_{\alpha,2}-\hat{\hat{d}}_2>0\ )=\lim_{n\rightarrow \infty}\mathbb{P}\left(\hat{p}_{\alpha}>0\right) =0. \label{quantile3}\end{equation}

After these preparations we are able to prove  the first part of Theorem \ref{thm2}, i.e. we show that the bootstrap test has asymptotic level $\alpha$ as specified in
  \eqref{test}.  It follows from \eqref{MLcons} that in the case
 $\hat{d}_2= d_2(\hat \beta_1, \hat \beta_2)\geq\epsilon_2$ the constrained estimators $\hat{\hat{\beta}}_1$ and $\hat{\hat{\beta}}_2$ coincide with the unconstrained OLS-estimators
 $\hat{\beta_1}$ and $\hat{\beta}_2$, respectively.   This yields in particular $\hat{\hat{d}}_2=d_2(\hat{\hat{\beta}}_1, \hat{\hat{\beta}}_2)=\hat{d}_2$ whenever $\hat d_2 \geq \varepsilon_2$.

If $d_2>\epsilon_2$ we   have
\begin{eqnarray*}
\mathbb{P}(\hat d_2<\hat{q}_{\alpha,2})
&=&\mathbb{P}(\hat d_2<\hat{q}_{\alpha,2},\ \hat d_2\geq\epsilon_2)+\mathbb{P}(\hat d_2<\hat{q}_{\alpha,2},\ \hat d_2<\epsilon_2) \\
&\leq &\mathbb{P} (\hat d_2<\hat{q}_{\alpha,2},\ \hat{\hat d}_2=\hat d_2 )+\mathbb{P}(\hat d_2<\epsilon_2)\\
&\leq &\mathbb{P} (\hat{\hat d}_2 <\hat{q}_{\alpha,2}  )+\mathbb{P}\Big(\frac{\sqrt{n}(\hat d_2-d_2)}{\sigma_{d_2}}<\frac{\sqrt{n}(\epsilon_2-d_2)}{\sigma_{d_2}}\Big).
\end{eqnarray*}
Observing that
$\epsilon_2-d_2<0$, it now follows from Theorem \ref{thm1} that  the second term is of order $o(1)$. On the other hand, we have from
$\eqref{quantile3}$ that the first term is of the same order, which gives
$\lim_{n_1,n_2\rightarrow \infty}\mathbb{P}(\hat{d_2}<\hat{q}_{\alpha,2})=0$
and proves the first part of Theorem \ref{thm2} in the case $d_2 > \epsilon_2$.

For a proof of the corresponding statement in the case  ${d}_2=\epsilon_2$   we note that it follows again from \eqref{quantile3}
{\begin{eqnarray} \label{decomp}
 \mathbb{P}(\hat d_2 < \hat{q}_{\alpha,2}) \nonumber
&=&\mathbb{P}(\hat d_2 <\hat{q}_{\alpha,2},\ \hat d_2\geq\epsilon_2)+\mathbb{P}(\hat d_2 <\hat{q}_{\alpha,2},\
\hat d_2<\epsilon_2)\\ \nonumber
&=&\mathbb{P} (\hat d_2 <\hat{q}_{\alpha,2},\ \hat{\hat d}_2=\hat d_2 )+\mathbb{P}(\hat d_2 <\hat{q}_{\alpha,2},\ \hat{\hat d }_2=\epsilon_2) - \mathbb{P}(\hat d_2 < \hat q_{\alpha,2}, \hat d_2 =  \epsilon_2) \\ \nonumber
&=&\mathbb{P} (\hat d_2 <\hat{q}_{\alpha,2},\ \hat{\hat d}_2=\hat d_2 )+\mathbb{P}(\hat d_2 <\hat{q}_{\alpha,2},\ \hat{\hat d }_2=\epsilon_2) + o(1)\\ \nonumber
& = &\mathbb{P}\Big(\frac{\sqrt{n}(\hat d_2- d_2)} {\sigma_{d_2}}<\frac{\sqrt{n}(\hat{q}_{\alpha,2}-\hat{\hat d}_2)}{\sigma_{d_2}},\hat{\hat d}_2 =\epsilon_2\Big) +o(1)   \\ \nonumber
&=&\mathbb{P}\Big(\frac{\sqrt{n}(\hat d_2-d_2)}{\sigma_{d_2}}<\frac{\sqrt{n}(\hat{q}_{\alpha,2}-\hat{\hat d}_2)}{\sigma_{d_2}}\Big) \\
&& ~~~~~
-\mathbb{P}\big(\hat d_2-d_2<\hat{q}_{\alpha,2}-\hat{\hat d}_2,\ \hat{\hat d}_2> \varepsilon_2 \big)+o(1) ,
  \end{eqnarray}
	where the third equality is a consequence of the fact that  $\sqrt{n} (\hat d_2 - d_2)$ is asymptotically normal distributed, which gives
	$\mathbb{P}(\hat d_2 < \hat q_{\alpha,2}, \hat d_2 = \epsilon_2) \leq\mathbb{P}(\hat d_2 = \epsilon_2) = o(1) .$}
  If ${\hat{\hat d}}_2 > \varepsilon_2$ it follows that $\hat d_2 = {\hat{\hat d}}_2 > \varepsilon_2=d_2$ and consequently the second term in \eqref{decomp} can be bounded by (observing again \eqref{quantile3})
  $\mathbb{P}(\hat d_2 - d_2 < \hat q_{\alpha,2} - {\hat{\hat d}}_2, \hat d_2  > \varepsilon_2) \leq
  \mathbb{P} (\hat q_{\alpha,2} - {\hat{\hat d}}_2 > 0) = o(1).$
  {Therefore we obtain from Theorem \ref{thm1}, $\eqref{quantile2}$ and \eqref{decomp} that
 $\lim_{n\rightarrow\infty}\mathbb{P}(\hat{d}_2<\hat{q}_{\alpha,2})=\Phi(u_\alpha)=\alpha,$}
which  completes
the proof of part (1) of Theorem \ref{thm2}.

\medskip

\textit{Proof of (2).}
Finally, we consider the case $d_2<\epsilon_2$ and show the consistency of the test \eqref{test}.    Theorem \ref{thm1} implies that $\hat d_2 {\stackrel{\mathbb{P} }{\longrightarrow}} d_2$. Since $d_2<\epsilon_2$, there exists a constant $\delta > 0$ such that $\mathbb{P}(\hat d_2 < \varepsilon_2-\delta) \to 1$. Hence the assertion will follow if we establish that $\mathbb{P}(\hat q_{\alpha,2} > \varepsilon_2-\delta) \to 1$. To show this, denote by $F_{b_1,b_2,s_1,s_2}^{(n)}$ the conditional distribution function of $\hat d^*_2$  given ${\hat{\hat \beta}}_\ell = b_\ell, \hat \sigma_\ell = s_\ell \ ( \ell=1,2)$. Since $\mathbb{P}(\max_{\ell=1,2}  | \hat \sigma_\ell - \sigma_\ell| \leq r) \to 1$ for any $r>0$, it suffices to establish that for some $r>0$
\[
\sup \big \{  F_{b_1,b_2,s_1,s_2}^{(n)}(\varepsilon_2 - \delta) \mid b_\ell \in B_\ell, \ell=1,2; \max_{\ell =1,2}  |s_\ell - \sigma_\ell|\leq r  \big \} \to 0.
\]
By uniform continuity of the map $(b_1,b_2) \mapsto d_2(b_1,b_2)$   it suffices to prove that for all $\eta>0$ and $\ell =1,2$
\begin{eqnarray}
\sup  \Big \{ \mathbb{P}(|\hat \beta_\ell^* - b_\ell| \geq \eta \mid \hat{\hat{\beta}}_k = b_k, \hat \sigma_k = s_k, k=1,2) \Big |  ~ {(b_\ell,s_\ell): |s_\ell - \sigma_\ell| \leq r, b_\ell \in B_\ell,\ell=1,2} \Big \} \to 0. \nonumber
\\
\label{eq:helpkons}
\end{eqnarray}
We will only prove the above statement for $\ell=1$ since the case $\ell=2$ follows by exactly the same arguments. For  $i=1,...,k_1, j=1,...,n_{1,i}$ let $e_{i,j}$ i.i.d. $\sim \mathcal{N}(0,1)$ and define
\begin{eqnarray*}
\psi_{a,1}^{(n)}(b) &:=& \sum_{i=1}^{k_1} \frac{n_{1,i}}{n_1} (m_1(x_{1,i},a) - m_1(x_{1,i},b))^2~,~
\hat \gamma_s :=  \frac{1}{n_1}\sum_{i=1}^{k_1}\sum_{j=1}^{n_{1,i}} (s e_{i,j})^2
\\
\hat \psi_{a,s}(b) &:=& \frac{1}{n_1}\sum_{i=1}^{k_1}\sum_{j=1}^{n_{1,i}} (m_1(x_{1,i},a) + se_{i,j} - m_1(x_{1,i},b))^2.
\end{eqnarray*}
By construction, the conditional distribution of $\hat \beta_1^*$ given $\hat{\hat{\beta}}_1 = a, \hat \sigma_1 = s$ is equal to the distribution of the random variable $\hat b_{a,s} := \arg\min_{b \in B_1} \hat \psi_{a,s}(b)$. On the other hand, $\arg\min_{b \in B_1} \hat \psi_{a,s}(b) = \arg\min_{b \in B_1} (\hat \psi_{a,s}(b) - \hat\gamma_s)$, and
\[
\hat \psi_{a,s}(b) - \hat\gamma_s = \psi_{a,1}^{(n)}(b) + 2s\sum_{i=1}^{k_1}(m_1(x_{1,i},a)- m_1(x_{1,i},b)) \frac{1}{n_{1}}\sum_{j=1}^{n_{1,i}}  e_{i,j}.
\]
{Observing that the terms $|m_1(x_{1,i},a)- m_1(x_{1,i},b)| $} are uniformly bounded (with respect to $a,b \in B_1$ and $ x_{1,i} \in {\cal X}$) it follows that
\[
R_n := \sup_{|s - \sigma_1| \leq r}\sup_{a,b \in B_1} \Big| 2s\sum_{i=1}^{k_1}(m_1(x_{1,i},a)- m_1(x_{1,i},b)) \frac{1}{n_{1}}\sum_{j=1}^{n_{1,i}}  e_{i,j}\Big| = o_\mathbb{P}(1)
\]
since $\max_{i=1,...,k_1}|\frac{1}{n_{1}}\sum_{j=1}^{n_{1,i}}  e_{i,j}| = o_\mathbb{P}(1)$. Now we obtain from Assumption \ref{2.5} that, for sufficiently large $n$,
\begin{eqnarray*}
\sup_{(a,s): |s - \sigma_1| \leq r}  \mathbb{P}(|\hat \beta_1^* - a| \geq \eta | \hat{\hat{\beta}}_1 = a, \hat \sigma_1 = s)
%\\
\leq
\sup_{(a,s): |s - \sigma_1| \leq r} \mathbb{P}(|\hat b_{a,s} - a| \geq \eta)
%\\
&\leq&
\mathbb{P}( R_n \geq v_{\eta}/4) = o(1).
\end{eqnarray*}
Thus \eqref{eq:helpkons}  follows, which completes the proof of Theorem \ref{thm2}.  \hfill $\Box$

\paragraph{Proof of Theorem \ref{thm3}:} \label{a.3}
Recall the definition of the estimator $\hat d_\infty$ in \eqref{dsup} and define the random variables
\begin{eqnarray} \label{Dn}
D_n &=& \sqrt{n} \ (\hat d_\infty - d_\infty) = \sqrt{n} \ \big( \max_{x \in \mathcal{X}} | m_1(x,\hat \beta_1) - m_2(x,\hat \beta_2)|-d_\infty\big), \\ \label{Zn}
Z_n &=& \sqrt{n} \ \big( \max_{x \in \mathcal{E}} | m_1(x,\hat \beta_1) - m_2(x,\hat \beta_2)|-d_\infty\big).
\end{eqnarray}
We will use similar arguments as given in \cite{raghavachari1973} and show  that
\begin{equation}\label{proofthm3a}
R_n = D_n - Z_n = o_{\mathbb{P}} (1),
\end{equation}
\begin{equation}\label{proofthm3b}
Z_n \dn \mathcal{Z},
\end{equation}
which proves the assertion of Theorem \ref{thm3}. For a proof of \eqref{proofthm3a}
we recall the definition of the "true" difference $\Delta (x,\beta_1,\beta_2)$ in \eqref{truediff}
and the
 definition of the process $\{p_n (x)\}_{x \in \mathcal{X}}$
in \eqref{delta}. It follows from \eqref{weakd} and the
  continuous mapping theorem that
\begin{equation} \label{p32}
\lim_{n_1,n_2 \to \infty} \mathbb{P} \big( \max_{x \in \mathcal{X}} |p_n(x)| > a_n\big) = 0
\end{equation}
as $n_1,n_2 \to \infty, \ n/n_1 \to \lambda \in (1,\infty)$, where $a_n= {\log n}/{\sqrt{n}}$.
By the representation $p_n(x) = G_n(x) + o_{\mathbb{P}}(n^{-1/2})$ uniformly in $x \in \mathcal{X}$ and the definition of $G_n$ in \eqref{delta} we have for every $\eta > 0$
\begin{equation} \label{p32a}
\lim_{\delta \downarrow 0} \lim_{n_1,n_2 \to \infty} \mathbb{P}\Big (\sup_{\|x-y\| < \delta} \sqrt{n} |p_n(x)- p_n(y)| > \eta \Big) = 0,
\end{equation}
where $\| \cdot \|$ denotes a norm on $\mathcal{X} \subset \mathbb{R}^d$.
In the following discussion define the sets
 {\begin{equation} \label{p33}
\mathcal{E}^{\mp}_n = \big \{ x \in \mathcal{X} \mid |\mp d_\infty - \Delta(x,\beta_1,\beta_2)| \leq a_n \big \}
\end{equation}}
and $ \mathcal{E}_n = \mathcal{E}^+_n \cup \mathcal{E}^-_n$, then it follows from the definition of $R_n$ and \eqref{p32} that
\begin{eqnarray*}
0 \leq R_n &=& \sqrt{n} \Big( \max_{x \in \mathcal{X}} |\Delta(x, \hat \beta_1, \hat \beta_2)| - \max_{x \in \mathcal{E}} |\Delta (x, \hat \beta_1, \hat \beta_2)| \Big) \\
&=& \sqrt{n} \Big( \max_{x \in \mathcal{E}_n} |\Delta(x, \hat \beta_1, \hat \beta_2)| - \max_{x \in \mathcal{E}} |\Delta (x, \hat \beta_1, \hat \beta_2)| \Big) + o_{\mathbb{P}}(1) \\
& \leq & \max (R^-_n, R^+_n) +  o_{\mathbb{P}}(1),
\end{eqnarray*}
where
% the quantities $R^-_n$ and $R^+_n$ are defined by
$
R^{\mp}_n = \sqrt{n} \big( \max_{x \in \mathcal{E}^{\mp}_n} |\Delta(x, \hat \beta_1, \hat \beta_2)| - \max_{x \in \mathcal{E}^{\mp}} |\Delta (x, \hat \beta_1, \hat \beta_2)|  \big).
$
We now prove the estimate $R^\mp_n = o_{\mathbb{P}}(1)$, which completes the proof of assertion \eqref{proofthm3a}. For this purpose we restrict ourselves to the random variable $R^+_n$ (the assertion for $R^-_n$ is obtained by similar arguments). Note that $\mathcal{E}^+  \subset \mathcal{E}^+_n$ and therefore it follows %, recalling the notation \eqref{test},
that
\begin{eqnarray}\label{p34}
0\leq R^+_n &=& \sqrt{n} \Big( \max_{x \in \mathcal{E}^{+}_n}  \Delta(x, \hat \beta_1, \hat \beta_2) - \max_{x \in \mathcal{E}^{+}}  \Delta (x, \hat \beta_1, \hat \beta_2 ) \Big) + o_{\mathbb{P}}(1) \\ \nonumber
& \leq & \max_{x \in \mathcal{E}^+_n} \sqrt{n} p_n(x) - \max_{x \in \mathcal{E}^+} \sqrt{n}p_n(x) + \sqrt{n}
\Big \{ \max_{x \in \mathcal{E}^+_n} \Delta(x,\beta_1,\beta_2) - d_\infty \Big \}+ o_{\mathbb{P}}(1) \\ \nonumber
&=& \max_{x \in \mathcal{E}^+_n} \sqrt{n} p_n(x) - \max_{x \in \mathcal{E}^+} \sqrt{n}p_n(x) + o_{\mathbb{P}}(1).
\end{eqnarray}
Now define for $\gamma > 0$ the set
$\mathcal{E}^+ (\gamma) = \{x \in \mathcal{X} \mid \exists \ y \in \mathcal{E}^+ \quad \mbox{with} \quad \| x-y\| < \gamma \}$
 and the constant
  {$\delta_n = 2 \inf \{ \gamma > 0 \mid \mathcal{E}^+_n \subset \mathcal{E}^+(\gamma) \}$}.
Obviously $\mathcal{E}^+_n \subset \mathcal{E}^+ (\delta_n)$ and the sequence $(\delta_n)_{n \in \mathbb{ N}} $ is decreasing, such that $\delta:= \lim_{n \to \infty} \delta_n$ exists. By the definition of $\delta_n$ we have $\mathcal{E}^+_n  \not\subset \mathcal{E}^+ ({\delta_n/4})$. Consequently, there exist $x_n \in \mathcal{E}^+_n \subset \mathcal{X}$ such that $\| x_n - y \| \geq {\delta_n/4}$ for all $y \in \mathcal{E}^+$ and all $n \in \en$. The sequence $(x_n)_{n \in \mathbb{N}}$ contains a convergent subsequence (because    $\mathcal{X}$ is compact), say $(x_{n_k})_{k \in \mathbb{N}}$ which satisfies
$
  \lim_{k \to \infty} x_{n_k} = x \in \mathcal{X},$ $
  d_\infty = \lim_{k \to \infty} \Delta(x_{n_k},\beta_1,\beta_2)=\Delta(x,\beta_1,\beta_2).
$
Consequently, $x \in \mathcal{E}^+$, but by construction $\| x_{n_k}-x \| \geq  {\delta_n/4}$ for all $k \in \mathbb{N}$, which is only possible if $\delta = \lim_{n \to \infty} \delta_n=0$.

Now it follows from inequality \eqref{p34} for the sequence $(\delta_n)_{n \in \mathbb{N}}$
\begin{eqnarray*}
o_{\mathbb{P}}(1) \leq R^+_n    &\leq&   \max_{x \in \mathcal{E}^+(\delta_n)} \sqrt{n} p_n (x) - \max_{x \in \mathcal{E}^+} \sqrt{n} p_n(x) + o_{\mathbb{P}}(1) \\
  &\leq &  \max_{\|y-x\|\leq \delta_n} \sqrt{n}|p_n(x) - p_n (y) | +o_{\mathbb{P}}(1) = o_{\mathbb{P}}(1),
\end{eqnarray*}
where the last estimate is a consequence of \eqref{p32a}. A similar statement for $R^-_n$ completes the proof of \eqref{proofthm3a}.

For a proof of the second assertion \eqref{proofthm3b} we define the random variable
$$
\tilde Z_n = \max \big \{ \max_{x \in \mathcal{E}^+} \sqrt{n} p_n(x) ; \ \max_{x \in \mathcal{E}^-}  (- \sqrt{n} p_n(x)) \big \},
$$
then it follows from \eqref{weakd} and the continuous mapping theorem that $\tilde Z_n \stackrel{\mathcal{D}}{\longrightarrow }\mathcal{Z}$, where the random variable $\mathcal{Z}$ is defined in \eqref{norm2}. Observing the uniform convergence in  \eqref{p32} we have as $n_1,n_2 \to \infty$
\begin{eqnarray*}
\mathbb{P}(Z_n \leq t) &=& \mathbb{P} \big(Z_n \leq t, \ \max_{x \in \mathcal{X}}|p_n(x)| < \tfrac {d_\infty}{2}\big) + o(1)
 =  \mathbb{P} \big( {\tilde Z_n \leq t} , \ \max_{x \in \mathcal{X}}|p_n(x)| < \tfrac {d_\infty}{2}\big)  + o(1)  \\
&=& \mathbb{P} (\tilde Z_n \leq t) + o (1) = \mathbb{P}(\mathcal{Z} \leq t) + o (1).
\end{eqnarray*}
This proves the remaining statement and completes the proof of Theorem \ref{thm3}.   \hfill $\Box$

\medskip

\paragraph{Proof of Theorem \ref{thm43}:}  \label{A4}
We begin by noting that the statement \eqref{consistence2} follows by exactly the same arguments as given in the proof of \eqref{consistence} in Theorem \ref{thm2} once we note that the mapping  $(b_1,b_2)\mapsto d_\infty(b_1,b_2)$ is uniformly continuous. The details are omitted for the sake of brevity. \\
Throughout the remaining proof, let $\hat{\hat{\beta_1}}$ and $\hat{\hat{\beta_2}}$ denote the  estimators defined by 
\[ 
{\hat{\hat{\beta}}_{\ell}}= \left\{
\begin{array} {ccc}
\hat \beta_\ell & \mbox{if} & \hat d_\infty \geq \varepsilon_\infty \\
\tilde \beta_\ell & \mbox{if} & \hat d_\infty < \varepsilon_\infty
\end{array}  \right. \quad \ell=1,2,
\]
where $\tilde \beta_1, \tilde \beta_2$ denote the OLS-estimators of the parameters $\beta_1, \beta_2$ under the constraint $d_\infty(\beta_1,\beta_2) = \epsilon_\infty$ 
and define
\begin{eqnarray*} \label{tay1}
p_n^* (x)  &:=&
\big(m_1(x,\hat\beta^*_1)-m_1 (x,\hat{\hat{\beta}}_1 )\big)
- \big(m_2(x,\hat\beta^*_2)-m_2 (x,\hat{\hat{\beta}}_2 )\big) , \\
\label{eq:gnstar}
 G_n^*(x)
&=& ( \tfrac{\partial }{\partial b_1}  m_1(x,b_1)\big|_{b_1 = \hat{ \hat{ \beta}}_1})^T   (\hat\beta_1^*-\hat{\hat{\beta_1}})-  
(\tfrac{\partial }{\partial b_2}  m_2(x,b_2)\big|_{b_2 =\hat {\hat {\beta}}_2})^T
 (\hat\beta^*_2-\hat{\hat{\beta}}_2) .
\end{eqnarray*}
Similarly to the proof of Theorem \ref{thm1}, it is possible to establish that
\begin{eqnarray}
\label{consconstr_infty}
{\hat{\hat{\beta}}_{\ell}} &\stackrel{\mathbb{P}}{\longrightarrow}& \beta_\ell \quad \ell=1,2, \quad \mbox{whenever} \quad d_\infty \geq \varepsilon_\infty, \\
 \left\{\sqrt{n}  p_n^*(x)\right\}_{x\in \mathcal{X}} &\stackrel{\mathcal{D}}{\longrightarrow} &\left\{G(x)\right\}_{x\in \mathcal{X}}
\label{repn} \\
 p_n^*(x) & = &G_n^*(x) + o_\mathbb{P}(n^{-1/2}) \label{repn2}
\end{eqnarray}
 uniformly with respect to  $x \in \mathcal{X}$, where $\left\{G(x)\right\}_{x\in \mathcal{X}}$ denotes the Gaussian process defined in \eqref{G}. Here, the weak convergence in \eqref{repn} holds conditionally on ${\cal Y} $ in probability as well as unconditionally.

From now on assume that the null hypothesis $d_\infty\geq\epsilon_\infty$ is satisfied, and define %$\mathcal{F}^*_n = \mathcal{F}^{+*}_n \cup \mathcal{F}^{-*}_n$, where
\begin{eqnarray*}
\mathcal{F}_n^{\mp*} &=& \big \{ x \in \mathcal{X}  \mid   m_1(x,\hat{\beta}_1^*)-m_2(x,\hat{ \beta}_2^*) = \mp \hat{ d}_\infty^* \big \},
\\
\Ec_n^{\mp} &=& \big \{ x \in \mathcal{X}  \mid   m_1(x,\hat{ \hat\beta}_1)-m_2(x,\hat{ \hat \beta}_2) = \mp \hat{ \hat d}_\infty \big \},
\end{eqnarray*}
where
\begin{eqnarray*}
\hat{ d}_\infty^*  &=& \sup_{x \in {\cal X}} |  m_1(x,\hat{\beta}_1^*)-m_2(x,\hat{ \beta}_2^*) | ~,~~ \hat{ \hat d}_\infty  = \sup_{x \in {\cal X}} | m_1(x,\hat{ \hat\beta}_1)-m_2(x,\hat{ \hat \beta}_2)|~.
\end{eqnarray*}
From \eqref{consconstr_infty} and the continuous mapping theorem we obtain the existence of a sequence $(\gamma_n)_{\in \en}$ such that $\gamma_n \to 0$ and
\begin{eqnarray} \label{bndef}
\sup_{x\in \mathcal{X}}|\Delta(x,\hat{ \hat{\beta}}_1,\hat{ \hat{\beta}}_2)-\Delta(x,\beta_1,\beta_2)|=o_{\mathbb{P}}(\gamma_n),\ \sup_{x\in \mathcal{X}}|\Delta(x,\hat \beta^*_1,\hat \beta^*_2)-\Delta(x,\hat{\hat{\beta}}_1,\hat{\hat{\beta}}_2)|=o_{\mathbb{P}}(a_n), \quad
\end{eqnarray}
where  $a_n= \log n / \sqrt{n}$ and the second statement follows from \eqref{repn}. Moreover, from the representation  \eqref{repn} we have for every $\eta > 0$
\begin{equation} \label{p32ab}
\lim_{\delta \downarrow 0} \limsup_{n_1,n_2 \to \infty} \mathbb{P}\Big (\sqrt{n} \sup_{\|x-y\| < \delta} |p^*_n(x)- p^*_n(y)| > \eta \Big) = 0.
\end{equation}

Now define $b_n= \max\{\gamma_n, a_n\}$ and consider the sets
$$
\mathcal{F}_n^\pm =\{ x \in {\cal X}~|~|\pm d_\infty  - \Delta (x,\beta_1,\beta_2)| \le b_n \}
$$
and $\mathcal{F}_n = \mathcal{F}_n^+ \cup \mathcal{F}_n^-$. Additionally, define the set
\[
\mathcal{E}^\pm (\gamma) = \{x \in \mathcal{X} \mid \exists \ y \in \mathcal{E}^\pm \quad \mbox{with} \quad \| x-y\| < \gamma \}
\]
for $\gamma>0$. At the end of the proof we shall show that there exists a sequence $\delta_n \to 0$ such that
\begin{equation} \label{eq:boot5help1}
\mathbb{P}\big(\mathcal{E}^{\pm}_n\cup\mathcal{F}^{\pm*}_n \subseteq \mathcal{F}_n^\pm \subseteq \mathcal{E}^\pm (\delta_n)\big) \to 1.
\end{equation}

In the special case $\# \mathcal{E} =1$ with $\Ec = \{x_0\}$ for some $x_0 \in \Xc$ we shall prove that additionally 
\begin{equation} \label{eq:boot5help2}
\sqrt{n}(\hat d_\infty^* - \hat{ \hat{ d}}_\infty)\stackrel{\mathcal{D}}{\rightarrow} G(x_0)  \quad \mbox{conditionally on ${\cal Y} $ in probability}.
\end{equation}

Given \eqref{eq:boot5help1} and \eqref{eq:boot5help2} we prove \eqref{level2} and \eqref{level2.1}. For a proof of \eqref{level2} note that $\hat{q}_{\alpha,\infty}$ is the $\alpha$-quantile of the bootstrap test statistics $\hat d_\infty^*$ conditionally on ${\cal Y} $. 
%Hence it holds that
%\[\alpha=\mathbb{P}\big(\hat d_\infty^*<\hat{q}_{\alpha,\infty}\big| {\cal Y} \big)=\mathbb{P}\big(\sqrt{n}(\hat d_\infty^*-\hat{\hat{d}}_\infty)<\sqrt{n}(\hat{q}_{\alpha,\infty}-\hat{\hat{d}}_\infty)\big| {\cal Y} \big),\]
%almost surely. 
Thus the $\alpha$-quantile of the distribution of $\sqrt{n}(\hat d_\infty^*-\hat{\hat{d}}_\infty)$ conditionally on ${\cal Y} $ is of the form
$\hat{p}_{\alpha,\infty}:=\sqrt{n}(\hat{q}_{\alpha,\infty}-\hat{\hat{d}}_\infty)$
and, by \eqref{eq:boot5help2}, satisfies
$\hat{p}_{\alpha,\infty}\stackrel{\mathbb{P}}\longrightarrow z_{\alpha}$,
where $z_{\alpha}$ denotes the $\alpha$-quantile of the distribution of $G(x_0)$. With
$\sigma^2_{d_\infty}:={\rm Var}(G(x_0))$,
  it now follows from {Lemma 21.2} in \cite{vaart1998}
${\hat{p}_{\alpha}}/{\sigma_{d_\infty}}\stackrel{\mathbb{P}}\longrightarrow u_{\alpha},$
where $u_{\alpha}$ denotes the $\alpha$-quantile of the standard normal distribution.
This result is the analogue of \eqref{quantile2}  in the proof of Theorem \ref{thm2},
and \eqref{level2} now follows by exactly the same arguments as given in the proof of~\eqref{level} in Theorem~\ref{thm2}.

Next, we derive a preliminary result that will be used to prove \eqref{level2.1}. 
%Let 
%\[
%\Zc_n^* := \sqrt{n}\max\big\{\max_{x \in \mathcal{F}^{+*}} p_n^*(x) , \max_{x \in \mathcal{F}^{-*}} (-p_n^*(x) )\big\}.
%\] 
%From \eqref{repn2} we obtain $\sqrt{n}(\hat d_\infty^*-\hat{\hat{d}}_\infty) = \Zc_n^* + o_\mathbb{P}(1)$. 
Define for the sequence $\delta_n$ from~\eqref{eq:boot5help1} 
\[
\widetilde{\Zc}^*_n := \sqrt{n}\max\big\{\max_{x \in \Ec^+(\delta_n)} p_n^*(x) , \max_{x \in \Ec^-(\delta_n)} (-p_n^*(x) )\big\}.
\]
From \eqref{eq:boot5help1} we obtain
\begin{eqnarray} \nonumber
\hat d_\infty^*-\hat{\hat{d}}_\infty &=& \max_{x\in \Xc} |\Delta(x,\hat\beta_1^*,\hat\beta_2^*)| - \hat{\hat{d}}_\infty
\\ 
&=&  \max \{\max_{x \in \Ec^+(\delta_n)} \Delta(x,\hat\beta_1^*,\hat\beta_2^*), \max_{x \in \Ec^-(\delta_n)} (-\Delta(x,\hat\beta_1^*,\hat\beta_2^*))\}  - \hat{\hat{d}}_\infty + o_{\mathbb{P}}(n^{-1/2}). \quad \quad \quad \label{dstard1}
%\\
%&\leq& \max \{\max_{x \in \Ec^+(\delta_n)} \Delta(x,\hat\beta_1^*,\hat\beta_2^*) - \Delta(x,\hat{\hat\beta}_1,\hat{\hat\beta}_2) , \max_{x \in \Ec^-(\delta_n)} (-\Delta(x,\hat\beta_1^*,\hat\beta_2^*))\}  + o_{\mathbb{P}}(1)
\end{eqnarray}
Moreover,
\[
\max_{x \in \Ec^\pm(\delta_n)} (\pm \Delta(x,\hat\beta_1^*,\hat\beta_2^*) - \hat{\hat d}_\infty) \leq \max_{x \in \Ec^\pm(\delta_n)} (\pm \Delta(x,\hat\beta_1^*,\hat\beta_2^*) \mp \Delta(x,\hat{\hat\beta}_1,\hat{\hat\beta}_2))
\]
and thus
\begin{equation} \label{bound1}
\sqrt{n}(\hat d_\infty^*-\hat{\hat{d}}_\infty) \leq \widetilde{\Zc}_n^* + o_{\mathbb{P}}(1).
\end{equation}

%Equation \eqref{eq:boot5help1} implies that $\Zc_n^* \leq \widetilde{\Zc}^*$ with probability tending to one, and hence $q_{\Zc_n^*,\alpha} \leq q_{\widetilde \Zc^*,\alpha}$ with probability tending to one. 
%Note that $\widetilde \Zc^* \stackrel{\mathcal{D}}{\rightarrow} \Zc$ conditional on ${\cal Y} $ in probability. Together with Assumption \ref{cont} this implies $q_{\widetilde\Zc^*,\alpha} \to q_{\Zc,\alpha}$ in probability. Since by assumption $q_{\Zc,\alpha} < 0$, it follows that $q_{\widetilde\Zc^*,\alpha} < 0$ with probability tending to one. 

Denoting by $\hat p_{\alpha,\infty}$ the $\alpha$-quantile of $\sqrt{n}(\hat d^*_\infty - \hat{\hat d}_\infty)$ conditional on the data we have $\hat q_{\alpha,\infty} = \hat{\hat d}_\infty + n^{-1/2} \hat p_{\alpha,\infty}$. 

Define $F_{n}^*$ as the distribution function of $\sqrt{n}(\hat d^*_\infty - \hat{\hat d}_\infty)$ conditional on the data, $F_n$ 
as the distribution function of $\widetilde{\Zc}_n^*$ conditional on the data and  $F_{\Zc}$  as 
the distribution function of $\Zc$. By the definition of $\Zc$ and the results in \cite{TS1976} the function $F_{\Zc}$ can have at most one jump, and this jump can only be located at its left support point. Since $F_{\Zc}$ is continuous at $q_{\Zc,\alpha}$ and $q_{\Zc,\alpha}<0$ there exists $t_0 < 0$ such that $F_{\Zc}(t_0) < \alpha$ and $F_{\Zc}$ is continuous on $[t_0,\infty)$. Since $\widetilde{\Zc}^*_n$ converges weakly to $\Zc$ conditionally on the data in probability it follows that $\sup_{t\geq t_0} |F_n(t) - F_{\Zc}(t)| = o_{\mathbb{P}}(1)$. From \eqref{bound1}   we have
$\mathbb{P}(F_{n}^*(t) \geq F_n(t - \epsilon)~~\forall t) \to 1$ for any $\epsilon > 0$,
and the   uniform continuity of $F_\Zc$ on $[t_0,\infty)$ yields 
$\mathbb{P}(F_n^*(t) \geq F_{\Zc}(t) - \epsilon~~\forall t \geq t_0 ) \to 1
 $
 for any $\epsilon > 0$.
  Let $q_{\Zc,\alpha+\delta}$ denote the $\alpha+\delta$ quantile of $F_{\Zc}$. For arbitrary $\delta > 0$ it follows that $q_{\Zc,\alpha+\delta} > t_0$ and thus 
\[
o(1) = \mathbb{P}( F_n^*(q_{\Zc,\alpha+\delta}) \geq F_{\Zc}(q_{\Zc,\alpha+\delta}) - \delta/3) = \mathbb{P}( F_n^*(q_{\Zc,\alpha+\delta}) \geq \alpha + 2\delta/3).
\] 
Thus, by definition of $\hat p_{\alpha,\infty}$, we have 
\begin{equation}\label{quantile3.2}
\mathbb{P} (\hat p_{\alpha,\infty} > q_{\mathcal{Z},\alpha+\delta}) = o(1) \quad \forall \delta >0.
\end{equation}
Specifically, choosing $\delta > 0$ with $q_{\mathcal{Z},\alpha+\delta} < 0$ we obtain
\begin{equation}\label{quantile3.1}
\mathbb{P} (\hat q_{\alpha,\infty} > {\hat{\hat d}}_\infty) = \mathbb{P} (\hat p_{\alpha,\infty} > 0) = o(1).
\end{equation}
Given \eqref{quantile3.2} and \eqref{quantile3.1} we are ready to prove \eqref{level2.1}. First consider the case  ${d}_\infty=\epsilon_\infty$ and note that it follows from \eqref{quantile3.1} that
{\begin{eqnarray} %\label{decomp2}
\mathbb{P}(\hat d_\infty < \hat{q}_{\alpha,\infty}) 
&=&\mathbb{P}(\hat d_\infty <\hat{q}_{\alpha,\infty},\ \hat{\hat d}_\infty = \epsilon_\infty)+\mathbb{P}(\hat d_\infty <\hat{q}_{\alpha,\infty}, \hat{\hat d}_\infty > \epsilon_\infty) \nonumber
\\ 
&=&\mathbb{P}(\hat d_\infty <\hat{q}_{\alpha,\infty},\ \hat{\hat d}_\infty = \epsilon_\infty)+\mathbb{P}(\hat{\hat d}_\infty <\hat{q}_{\alpha,\infty}, \hat{\hat d}_\infty > \epsilon_\infty) \nonumber
%\\ 
%&=& \mathbb{P}(\hat d_\infty <\hat{q}_{\alpha,\infty}, \hat d_\infty<\epsilon_\infty)  + o(1) \nonumber \\ \nonumber
%&=&\mathbb{P} (\hat d_\infty <\hat{q}_{\alpha,\infty},\ \hat{\hat d}_\infty=\hat d_\infty )+\mathbb{P}(\hat d_\infty <\hat{q}_{\alpha,\infty},\ \hat{\hat d }_\infty=\epsilon_\infty) + o(1)
\\ \nonumber
& = &\mathbb{P}\Big(\sqrt{n}(\hat d_\infty- d_\infty)<\sqrt{n}(\hat{q}_{\alpha,\infty}-\hat{\hat d}_\infty),\hat{\hat d}_\infty =\epsilon_\infty\Big) +o(1)   
\\ \nonumber
&\leq&\mathbb{P}\Big(\sqrt{n}(\hat d_\infty-d_\infty)<\sqrt{n}(\hat{q}_{\alpha,\infty}-\hat{\hat d}_\infty)\Big) +o(1) 
\\
& = &  \mathbb{P}\Big(\sqrt{n}(\hat d_\infty-d_\infty) <  \hat p_{\alpha,\infty} \Big) +o(1)  ,\nonumber
%\\ \nonumber
%&\leq&\mathbb{P}\Big(\sqrt{n}(\hat d_\infty-d_\infty)<\sqrt{n}(\hat{q}_{\alpha,\infty}-\hat{\hat d}_\infty)\Big) +o(1)
%\\
%&& ~~~~~
%-\mathbb{P}\big(\hat d_\infty-d_\infty<\hat{q}_{\alpha,\infty}-\hat{\hat d}_\infty,\ \hat{\hat d}_\infty> \epsilon_\infty \big)+o(1) ,
\end{eqnarray}
where the second equality follows since on the event $\hat{\hat d}_\infty > \epsilon_\infty$ we have $\hat{\hat d}_\infty = {\hat d}_\infty $ and the third equality follows from \eqref{quantile3.1}. From \eqref{quantile3.2} we obtain for any $\delta > 0$
\[
\mathbb{P}\Big(\sqrt{n}(\hat d_\infty-d_\infty) <  \hat p_{\alpha,\infty} \Big) \leq \mathbb{P}\Big(\sqrt{n}(\hat d_\infty-d_\infty) < q_{\Zc,\alpha+\delta} \Big) + o(1) \to \alpha + \delta
\]
because  the distribution of $\Zc$ is continuous at $q_{\Zc,\alpha+\delta}$. Since $\delta > 0$ was arbitrary \eqref{level2.1} follows in the case ${d}_\infty=\epsilon_\infty$. 

\medskip

Next consider the case ${d}_\infty > \epsilon_\infty$. We have
\begin{eqnarray*}
\mathbb{P}(\hat d_\infty<\hat{q}_{\alpha,\infty})
&=&\mathbb{P}(\hat d_\infty<\hat{q}_{\alpha,\infty},\ \hat d_\infty\geq\epsilon_\infty)+\mathbb{P}(\hat d_\infty<\hat{q}_{\alpha,\infty},\ \hat d_\infty<\epsilon_\infty) \\
&\leq &\mathbb{P} (\hat d_\infty<\hat{q}_{\alpha,\infty},\ \hat{\hat d}_\infty=\hat d_\infty )+\mathbb{P}(\hat d_\infty<\epsilon_\infty)\\
&\leq &\mathbb{P} (\hat{\hat d}_\infty <\hat{q}_{\alpha,\infty}  )+\mathbb{P}(\sqrt{n}(\hat d_\infty-d_\infty)<\sqrt{n}(\epsilon_\infty-d_\infty)) = o(1)
\end{eqnarray*}
where the first term in the last line is of order $o(1)$ by \eqref{quantile3.1} and the second term vanishes since $\sqrt{n}(\epsilon_\infty-d_\infty) \to \infty$ while $|\sqrt{n}(\hat d_\infty-d_\infty)|$ converges weakly and thus is of order $O_{\mathbb{P}}(1)$. This completes the proof of \eqref{level2.1}.
\medskip

%\newpage

It remains to establish \eqref{eq:boot5help1} and \eqref{eq:boot5help2}. We begin with a proof of \eqref{eq:boot5help1}. Without loss of generality, we only prove the existence of $\delta_n \to 0$ with $\mathbb{P}(\Ec^+_n\cup\mathcal{F}^{+*}_n \subseteq \mathcal{F}_n^+ \subseteq \mathcal{E}^+ (\delta_n)) \to 1.$ We may assume that $\Ec^+ \neq \emptyset$, otherwise $\mathcal{E}^+ (\delta_n) $ is empty and it is straightforward to show that $\mathcal{F}^{+*}_n ,\mathcal{F}_n^+, \Ec^+_n$ will be empty with probability converging to one.  
Define $\delta_n = 2\cdot \inf \{ \gamma > 0 \mid \mathcal{F}^{+}_n \subset \mathcal{E}^+(\gamma) \}$.
Obviously $\Ec^+ \subset \Ec^+(\delta_n)$ provided that $\delta_n > 0$. Moreover, without loss of generality we assume that the sequence $b_n$ is non-increasing. As a consequence $(\delta_n)_{n \in \mathbb{ N}} $ is also non-increasing, such that $\delta:= \lim_{n \to \infty} \delta_n$ exists. By the definition of $\delta_n$ we have $\mathcal{F}^{+}_n\not\subset \Ec^+(\delta_n/4)$ unless $
\delta_n = 0$ in which case $\delta = 0$. Consequently, for each $n$ with $\delta_n > 0$ there exists an $x_n \in \mathcal{F}^{+}_n$ such that $\| x_n - x_0 \| \geq \delta_n/4$ for all $x_0 \in \Ec^+$. As  $\mathcal{X}$ is compact, there exists a convergent sub-sequence, say $(x_{n_k})_{k \in \mathbb{N}}$ with limit $\lim_{k \to \infty} x_{n_k} = x \in \mathcal{X}$ and
\[
d_\infty = \lim_{n_1,n_2\rightarrow\infty}\Delta(x_{n_k},\beta_1,\beta_2)=\Delta(x,\beta_1,\beta_2).
\]
Consequently, $x\in \Ec^+$, and from $\| x_{n_k}-x_0 \| \geq \delta/4$ for all $k \in \mathbb{N}, x_0 \in \Ec^+$ we obtain $\delta = \lim_{n \to \infty} \delta_n=0$.

Next, note that by \eqref{bndef}
\[
\sup_{x\in \Xc} |m_1(x,\hat{\beta}_1^*) - m_2(x,\hat{ \beta}_2^*) - (m_1(x,{\beta}_1) - m_2(x,{ \beta}_2)) | = o_{\mathbb{P}}(b_n)
\]
Hence 
\[
\mathbb{P}(\mathcal{F}^{+*}_n \subseteq \mathcal{F}_n^+) \leq \mathbb{P}\Big(\sup_{x\in \Xc} |m_1(x,\hat{\beta}_1^*) - m_2(x,\hat{ \beta}_2^*) - (m_1(x,{\beta}_1) - m_2(x,{ \beta}_2)) | \leq b_n/8 \Big) \to 1.
\]
The convergence $\mathbb{P}(\mathcal{E}^{+}_n \subseteq \mathcal{F}_n^+) \to 1$ follows by similar arguments, which
 establishes \eqref{eq:boot5help1}. 

\medskip
%\newpage

Finally, it remains to prove \eqref{eq:boot5help2} for the special case that $\Ec = \{x_0\}$. Assume without loss of generality that $\Ec = \Ec^+$ and $\Ec^-$ is empty. This implies that $\mathcal{F}^{-}_n, \Ec_n^-$ will be empty with probability converging  to one and $\mathcal{F}^{+}_n$ will contain at least one point with probability tending to one. Together with \eqref{dstard1} we obtain
\[
D_n^* := \sqrt{n}(\hat d_\infty^*-\hat{\hat{d}}_\infty) =  \sqrt{n}(\max_{x \in \Ec^+(\delta_n)} \Delta(x,\hat\beta_1^*,\hat\beta_2^*)-\hat{\hat{d}}_\infty) + o_{\mathbb{P}}(1),
% \max_{x\in \Xc} |\Delta(x,\hat{\hat\beta}_1,\hat{\hat\beta}_2)| + o_{\mathbb{P}}(1).
\]
%Defining the random variable
%\begin{eqnarray*}
%M_n^* &=& \sqrt{n}(\max_{x \in \mathcal{F}^{+*}} \Delta(x,\hat\beta_1^*,\hat\beta_2^*)-\hat{\hat{d}}_\infty),
%\end{eqnarray*}
and thus 
\begin{eqnarray*}
o_{\mathbb{P}}(1) &\leq& D_n^* - \sqrt{n}(\max_{x \in \Ec_n^+} \Delta(x,\hat\beta_1^*,\hat\beta_2^*)-\hat{\hat{d}}_\infty)
\\
&=& \sqrt{n} \Big( \max_{x \in \Ec^+(\delta_n)} \Delta(x, \hat \beta^*_1, \hat \beta^*_2) - \max_{x \in \Ec_n^+} \Delta(x, \hat \beta^*_1, \hat \beta^*_2)  \Big) + o_{\mathbb{P}}(1)  
\\ 
& \leq & \max_{x \in \Ec^+(\delta_n)} \sqrt{n} p_n^*(x) - \max_{x \in \Ec_n^+} \sqrt{n} p_n^*(x) + \sqrt{n}
\Big \{ \max_{x \in \Ec^+(\delta_n)} \Delta(x,\hat{ \hat{\beta}}_1,\hat{ \hat{\beta}}_2) - \hat{ \hat{ d}}_\infty \Big \}+ o_{\mathbb{P}}(1) 
\\ 
&=& \max_{x \in \Ec^+(\delta_n)}\sqrt{n}  p_n^*(x) - \max_{x \in \Ec_n^+} \sqrt{n} p_n^*(x)+ o_{\mathbb{P}}(1)
\\
&=& o_{\mathbb{P}}(1)
\end{eqnarray*}
where the last equality follows from \eqref{p32ab} and the second-to last equality 
is a consequence of \eqref{eq:boot5help1}. Thus
\[
\sqrt{n}(\hat d_\infty^*-\hat{\hat{d}}_\infty) = \sqrt{n}\Big (\max_{x \in \Ec_n^+} \Delta(x,\hat\beta_1^*,\hat\beta_2^* )-\hat{\hat{d}}_\infty\Big) + o_{\mathbb{P}}(1) = \sqrt{n} \max_{x \in \Ec_n^+} p_n^*(x) + o_{\mathbb{P}}(1) = \sqrt{n}p_n^*(x_0) + o_{\mathbb{P}}(1),
\]
where the last equality follows from a combination of \eqref{p32ab} and \eqref{eq:boot5help1}. Since $\sqrt{n}p_n^*(x_0)$ converges weakly to $G(x_0)$ conditionally on the data in probability the statement \eqref{eq:boot5help2} follows. This completes the proof of Theorem \ref{thm43}. 
\hfill $\Box$

\end{document}